\documentclass[letterpaper, aps, showpacs, onecolumn]{revtex4}

\usepackage{amsmath}
\usepackage{amssymb}
\usepackage{latexsym}
\usepackage{graphicx,epstopdf}
\usepackage{hyperref}

\begin{document}

\title{Lagrangian dynamics in inhomogeneous and thermal environments\\
An application of the Onsager-Machlup theory I}
\author{Alexander Jurisch}
\affiliation{ajurisch@ymail.com, Munich, Germany}

\begin{abstract}
We straight-forwardly derive the Onsager-Machlup Lagrangian from the Fokker-Planck equation and show that friction and dissipation are a natural property of the equation of motion. We develop a method to calculate the local variance $\sigma_{2}\,b(q)^{2}$ and identify this function as a Helmholtz-factor. In both meanings the function $b(q)$ describes properties of the environment. For application, we examine the free fall through a barometric medium and model a blow of wind by a solitonic pulse running through the medium. We treat harmonic oscillators immersed in a thermal bath, finding intuitive as well as counter-intuitive phenomena of friction. By allowing the temperature to be time-dependent, the dynamical process of cooling and heating becomes self-consistently available. We find a state of dynamical balance between system and environment. Last, we show that dynamical balance is related to adiabatic thermodynamic processes. In a special case, dynamical balance can induce a real phase-transition.
\end{abstract}
\maketitle
\section{Introduction}
A major drawback of the Onsager-Machlup theory is the determination of the functions, which describe the properties of the environment. These functions, $\{a,\,b\}$, are usually related to the  Langevin-equation
\begin{equation}
dq(t)\,=\,a[q(t)]\,dt\,+\,b[q(t)]\,dW(t)\quad.\nonumber
\end{equation}
The function $a[q(t)]$ describes a drift-field, the function $b[q(t)]$ describes properties of the disorder. The function $a[q(t)]$ is mostly easier to chose, be it ordinary Brownian motion, the Ornstein-Uhlenbeck process, the Cox-Ingersoll-Ross process \cite{Cox} or chemical reactions, but the function $b[q(t)]$ is problematic. Well-known choices are the cases of geometric Brownian motion $b[q(t)]\,=\,q(t)$ and $b[q(t)]\,=\,\sqrt{q(t)}$ for the Cox-Ingersoll-Ross process. However, these choices are of interest in financial physics, but for real physical systems the environmental function $b[q(t)]$ is usually unknown. Below, we suggest an easy and physically reasonable method, which allows the calculation of the function $b[q(t)]$ for any environment.

The Fokker-Planck equation is a consequence of the second order Kramers-Moyal expansion of the Boltzmann-equation, and only It\^o's lemma relates the Fokker-Planck equation to the Langevin-equation. While master-equations pursue a more microscopic point of view, the Fokker-Planck equation is a general diffusion-equation for macroscopic processes. Furthermore, the Fokker-Planck equation and the Onsager-Machlup Lagrangian are related by a transition-amplitude. This connection alone makes the Onsager-Machlup Lagrangian into the very candidate for a general and systematic approach to friction and dissipation.

Attempts to go beyond the usual Lagrangian are subject to severe restrictions, which are imposed by the Helmholtz-conditions \cite{Helmholtz}, see e.g. Nigam and Banerjee \cite{Nigam}, and Ostrogradsky's theorem \cite{Ostrogradsky}, see e.g. Woodard \cite{Woodard}. Ostrogradsky's theorem rules out Lagrangians with higher than first order derivatives, the Helmholtz-conditions provide the most general form of allowed Lagrangians within the frame of Ostrogradsky's theorem. The Onsager-Machlup Lagrangian fits into the frame of Ostrogradsky's theorem and the Helmholtz-conditions and, as a fact, the environmental function $b[q(t)]$ indeed has the meaning of a Helmholtz-factor. Mathematically, the Helmholtz-factor is a Jacobi-multiplier and works like Euler's integrating factor. Jacobi's fundamental work on Lagrangians \cite{Jacobi1, Jacobi2, Jacobi3, Jacobi4} is only rarely known, for a survey and further work on this topic see Nucci and Leach \cite{Nucci}, and references therein. We emphasize that if the Onsager-Machlup theory would not fit into the framework of Helmholtz and Ostrogradsky it would be meaningless, and no argument from stochastic calculus or statistical mechanics could reanimate this approach. Concerning Lagrangians, there are no stricter and more general conditions than the Helmholtz-conditions and Ostrogradsky's theorem. However, in their seminal paper Onsager and Machlup \cite{Machlup} proposed a second order variational-principle, which violates Ostrogradsky's theorem. In a sequel to this paper we elucidate this mistake and suggest a solution.

In order to describe dissipation and friction Bateman \cite{Bateman} invented a dual non-standard Lagrangian. Non-standard Lagrangians appear to be somewhat artificial, but as long as they obey Jacobi, Helmholtz and Ostrogradsky, they may be regarded as existent. A consequence of Bateman's Lagrangian is the Hamiltonian of Caldirola \cite{Caldirola} and Kanai \cite{Kanai}, which in fact is a Legendre-transform of an Onsager-Machlup Lagrangian. However, except it's relation to the Caldirola-Kanai Hamiltonian, Bateman's Lagrangian itself but does not exhibit any obvious connection or similarity to an Onsager-Machlup Lagrangian. Bateman's Lagrangian is still of interest, for recent work see e.g. Galley \cite{Galley} and Martínez-Pérez et. al. \cite{Martinez}. For an exhaustive review on Bateman, Caldirola-Kanai and related topics see Dekker \cite{Dekker5}.

For completeness, we also shall mention the Lagrangian approach by Martin, Siggia and Rose \cite{Martin}, de Dominici \cite{deDominici}, and Janssen \cite{Janssen}, which, with respect to the dual variables, is formally more closely related to Bateman's Lagrangian than to the Onsager-Machlup Lagrangian.

Employing the Helmholtz-conditions and Jacobi's method Havas \cite{Havas} initiated the search for even more exotic non-standard Lagrangians.

From our short survey, we can draw the following conclusion. On the Lagrangian way of physics many different approaches to friction and dissipation are to be found without any too obvious systematics. The only constraints are the Helmholtz-conditions and Ostrogradsky's theorem. It appears to be extremely difficult to built up a systematic Lagrangian generalization, which describes dissipation and friction in a natural manner. In this context, we also refer to the cumbersome procedure reported in Landau-Lifshitz \cite{Landau}, which is applicable only for Stokes-friction and highly problematic for Newtonian friction.

The work we present here shows that the Onsager-Machlup Lagrangian provides a natural approach to friction and dissipation on an utmost elementary and general level.
\newline

Our paper is organized as follows. First, we derive the Onsager-Machlup Lagrangian by exploiting the operator-ordering of the Fokker-Planck equation, which proves to be a much easier and straight-forward route than the application of stochastic calculus or covariance-arguments in curved spaces. Second, we give arguments for the inclusion of several types of interaction and discuss the equation of motion. Three types of forces are to be found: a force related to the interaction, a force purely generated by the properties of the environment, and a dissipative force, which describes Stokes-friction and Newtonian friction in a general way. It will turn out that the environmental function $b[q(t)]$ describes the properties of friction.

For a first application, we start with the phenomenological equation of the free fall with Newtonian friction. The same equation can be derived from the Onsager-Machlup Lagrangian, and by comparison this allows us to calculate the properties of the barometric medium given by the environmental function $b[q(t)]$. Knowing the properties of the environment, we can easily model a blow of wind by a solitonic pulse.

By the same argument as in the case of the barometric medium, we derive the environmental function, that applies for a bath of oscillators. We calculate the effects of the dissipative and environmental forces, and by allowing the thermal energy to be time-dependent, we observe the relaxation towards a state of dynamical balance between the system and the environment in a self-consistent way. Dynamical balance thereby is a state, where the motion evolves apparently without friction, since the effects of friction and the time-evolution of the thermal energy buffer each other. The state of dynamical balance is no equilibrium in a strict sense, because the system under consideration is not a many-body system, but very well may serve as a model for the behaviour of a many-body system.

Furthermore, we treat the case of two harmonically coupled oscillators in different scenarios of their environments. This yields three major results. First, we encounter that the attempt of a coupled system to gain dynamical balance in an inert thermal environment can behave counter-intuitively. Second, for a reactive thermal environment, we find behaviour, which may be interpreted as a real phase-transition. Third, for coupled oscillators immersed in environments at different thermal energy, we discover that already an apparently negligible perturbation due to the coupling can lead to essentially modified behaviour compared to the uncoupled case.

In the last section, we discuss the production of heat and entropy with respect to our results about the harmonic oscillator. This allows us to identify adiabatic thermodynamic processes with the state of dynamical balance. From this we may draw the conclusion that dynamical balance indeed is an analogon to the thermodynamic equilibrium of many-body systems.

\section{Derivation of the Lagrangian and the equation of motion}
We consider a general Brownian process in one dimension. The process is defined by the Langevin-equation
\begin{equation}
dq(t)\,=\,a[q(t)]\,dt\,+\,\sqrt{\sigma_{2}}\,b[q(t)]\,dW(t)\quad,
\label{General1}\end{equation}
where $a[q(t)]$ is a drift-field, $\sigma_{2}$ is the strength of the variance, $b[q(t)]$ is a characteristic function of the environment, and $dW(t)$ is a Gaussian Wiener-process. Note that for the function $b[q(t)]\neq 0$ always holds. If the environment has no specifications in terms of $b[q(t)]$ then $b[q(t)]=1$.

It\^o's lemma relates the Langevin-equation to a transition-amplitide $G(q,\,t|q_{0},\,t_{0})$, given by the Fokker-Planck-equation
\begin{equation}
\partial_{t}\,G(q,\,t|q_{0},\,t_{0})\,=\,-\,\partial_{q}\left(a[q(t)]\,G(q,\,t|q_{0},\,t_{0})\right)\,+\,\frac{\sigma_{2}}{2}\,\partial_{q}^{2}\left(b[q(t)]^{2}\,G(q,\,t|q_{0},\,t_{0})\right)\quad.
\label{General2}\end{equation}
We can cast Eq. (\ref{General2}) into
\begin{equation}
\partial_{t}\,G(q,\,t|q_{0},\,t_{0})\,=\,-\,\hat{H}\,G(q,\,t|q_{0},\,t_{0})\quad,
\label{General3}\end{equation}
where the Hamiltonian is given by
\begin{equation}
\hat{H}\,=\,\partial_{q}\,a[q(t)]\,-\,\frac{\sigma_{2}}{2}\,\partial_{q}^{2}\,b[q(t)]^{2}\quad.
\label{General4}\end{equation}
The effects of operator-ordering can be included by noticing that the Hamiltonian in Eq. (\ref{General4}) can be expanded into
\begin{equation}
\hat{H}\,=\,\partial_{q}\,a[q(t)]\,+\,a[q(t)]\,\partial_{q}\,-\,\sigma_{2}\,\left(\left(\partial_{q}\,b[q(t)]\right)^{2}\,+\,b[q(t)]\,\partial_{q}^{2}\,b[q(t)]\right)\,-\,2\,\sigma_{2}\,b[q(t)]\,\partial_{q}\,b[q(t)]\,\partial_{q}\,-\,\frac{\sigma_{2}}{2}\,b[q(t)]^{2}\,\partial_{q}^{2}\quad,
\label{General5}\end{equation}
which corresponds to a fully differentiated Fokker-Planck equation.
By making use of the correspondence-principle, $\hat{p}=-\,i\,\partial_{q}$, we arrive at the Hamiltonian
\begin{equation}
\mathcal{H}(p,\,q)\,=\,\partial_{q}\,a[q(t)]\,-\,\sigma_{2}\,\left(\left(\partial_{q}\,b[q(t)]\right)^{2}\,+\,b[q(t)]\,\partial_{q}^{2}\,b[q(t)]\right)\,+\,i\,\left(a[q(t)]\,-\,2\,\sigma_{2}\,b[q(t)]\,\partial_{q}\,b[q(t)]\right)\,p\,+\,\frac{\sigma_{2}}{2}\,b[q(t)]^{2}\,p^{2}\,.
\label{General6}\end{equation}
The Onsager-Machlup Lagrangian can now be derived by the path-integral. The transition-amplitude $G(q,\,t|q_{0},\,t_{0})$ is given by
\begin{eqnarray}
G(q,\,t|q_{0},\,t_{0})&=&\int\,\mathcal{D}[q(t)]\,\mathcal{D}[p(t)]\,\exp\left[\int_{t_{0}}^{t}d\tau\,\left(i\,p(\tau)\,\dot{q}(\tau)\,-\,\mathcal{H}(p,\,q)\right)\right]\nonumber\\
&=&\int\,\mathcal{D}[q(t)]\,\int_{-\infty}^{\infty}\frac{dp}{2\pi}\,\exp\left[\int^{t}_{t_{0}}d\tau\,\left(i\,p\,\dot{q}(\tau)\,-\,\mathcal{H}(p,\,q)\right)\right]\nonumber\\
&=&\int\,\mathcal{D}[q(t)]\,\frac{1}{\sqrt{2\,\pi\,\sigma_{2}\,\int^{t}_{t_{0}}d\tau\,b[q(\tau)]^{2}}}\,\exp\left[-\,\int^{t}_{t_{0}}d\tau\,\mathcal{L}^{*}(\dot{q},\,q)\right]\nonumber\\
&=&\int\,\mathcal{D}[q(t)]\,\mathcal{N}[q(t)]\,\exp\left[-\,\int^{t}_{t_{0}}d\tau\,\mathcal{L}^{*}(\dot{q},\,q)\right]\quad.
\label{General7}\end{eqnarray}
The Lagrangian in Eq. (\ref{General7}) is the Onsager-Machlup Lagrangian, and is given by
\begin{equation}
\mathcal{L}^{*}(\dot{q},\,q)\,=\,\frac{1}{2\,\sigma_{2}}\,\left(\frac{\dot{q}(t)\,-\,a[q(t)]\,+\,2\,\sigma_{2}\,b[q(t)]\,\partial_{q}\,b[q(t)]}{b[q(t)]}\right)^{2}\,+\,\partial_{q}\,a[q(t)]\,-\,\sigma_{2}\,\left(\left(\partial_{q}\,b[q(t)]\right)^{2}\,+\,b[q(t)]\,\partial_{q}^{2}\,b[q(t)]\right)\,.
\label{General8}\end{equation}
Thus, the Fokker-Planck equation and the Onsager-Machlup Lagrangian are directly and straight-forwardly related to each other in a natural way. Consequently, it is irrelevant which type of description is chosen to calculate propagation. The Onsager-Machlup Lagrangian is useful to examine the dynamics of particles on their most probable path through a disordered environment, while the dynamics of a field should better be treated by the Fokker-Planck equation instead of using the transition-amplitude Eq. (\ref{General7}). The application of Trotter's formula thereby provides a much better access to the dynamics than cumbersome discretization-techniques. This, by one stroke, also rules out any problems, which appear to be involved with the transition-amplitude and the Onsager-Machlup Lagrangian, as it is suspected. These problems are fundamental in nature, and we shall address them now.

\subsection{A note on covariance}
Much effort has been put into the question of how the Onsager-Machlup Lagrangian must look like in order to be invariant under coordinate-transform. This either means curved spaces, as well as stochastic coordinate-transform. The results of this work differ from our present result, and so we shall give a short discussion and formulate an objection.
Our objection thereby relies on the clear and straight-forward hierarchy Fokker-Planck equation $\rightarrow$ Onsager-Machlup Lagrangian, as we have demonstrated it. Thus, there is a sound physical foundation.

\subsubsection{Covariance in curved spaces}
Many attempts have been made to construct an Onsager-Machlup Lagrangian, that is covariant under general coordinate-transform in curved spaces. Most noticeable amongst them are e.g. Dekker \cite{Dekker1, Dekker2, Dekker3, Dekker4} and Graham \cite{Graham1, Graham2}. Starting-point of these examinations is the incomplete Onsager-Machlup Lagrangian as originally proposed by Onsager and Machlup \cite{Machlup, Onsager}
\begin{equation}
\mathcal{L}^{*}(\dot{q},\,q)\,=\,\frac{1}{2\,\sigma_{2}}\,\left(\frac{\dot{q}(t)\,-\,a[q(t)]}{b[q(t)]}\right)^{2}\quad.\nonumber
\end{equation}
The additional terms we have found above, see Eq. (\ref{General8}), are missing. The goal of all attempts to generate covariance was that the completed Onsager-Machlup Lagrangian shall meet with the Fokker-Planck equation, thus one tried the direction Onsager-Machlup Lagrangian $\rightarrow$ Fokker-Planck equation. In these attempts, the cart has been put before the horse.

To generate covariance, the function $b[q(t)]^{2}$, which in higher dimensions is a local the diffusion-matrix $D_{ij}$ was interpreted as a metrical tensor, and has been related to the Ricci-tensor $R_{ij}$. In doing so, additional terms have been calculated, that to some extend meet the terms we have calculated in Eq. (\ref{General8}). As far as we know, only Dekker \cite{Dekker1} has derived a result, that meets ours.

\subsubsection{Covariance and the stochastic Leibniz-rule}
On the context of stochastic calculus also many attempts have been made to derive an Onsager-Machlup Lagrangian, which meets the requirements of stochastic coordinate-transform. In stochastic coordinate-transform also the second order of the Leibniz-rule must be taken account, because the Wiener-differential cannot soundly be treated as a differential of first order. The Wiener-differential has a dimension $dW(t)\sim\sqrt{dt}$, such that the second order is needed to generate a sound differential $dt$. This odd connection is the base of Itô's formula.

The first attempt here has been made by Stratonovich \cite{Stratonovich2}, then followed by e.g. D\"urr and Bach \cite{Durr}, Horsthemke and Bach \cite{Horsthemke}, Itami and Sasa \cite{Itami}, Lau and Lubensky \cite{Lau}. Only recently Cugliandolo et. al. \cite{Cugliandolo1, Cugliandolo2} presented an exhaustive analysis about all aspects of a derivation of the Onsager-Machlup Lagrangian based on stochastic calculus.

The reduced version of the Onsager-Machlup Lagrangian, that leaves the transition-amplitude Eq. (\ref{General7}) invariant under stochastic coordinate-transform reads
\begin{equation}
\mathcal{L}^{*}(\dot{q},\,q)\,=\,\frac{1}{2\,\sigma_{2}}\,\left(\frac{\dot{q}(t)\,-\,a[q(t)]\,+\,\sigma_{2}\,b[q(t)]\,\partial_{q}\,b[q(t)]}{b[q(t)]}\right)^{2}\,+\,\partial_{q}\,a[q(t)]\quad.\nonumber
\end{equation}
This Lagrangian is also the result of some curved space covariance-treatments. Again it lacks the third term we have calculated in our version Eq. (\ref{General8}). However, by introducing
\begin{equation}
a[q(t)]\,=\,a_{1}[q(t)]\,+\,\sigma_{2}\,b[q(t)]\,\partial_{q}\,b[q(t)]\quad,\nonumber
\end{equation}
and inserting this into the Onsager-Machlup Lagrangian Eq. (\ref{General8}) one obtains
\begin{equation}
\mathcal{L}^{*}(\dot{q},\,q)\,=\,\frac{1}{2\,\sigma_{2}}\,\left(\frac{\dot{q}(t)\,-\,a_{1}[q(t)]\,+\,\sigma_{2}\,b[q(t)]\,\partial_{q}\,b[q(t)]}{b[q(t)]}\right)^{2}\,+\,\partial_{q}\,a_{1}[q(t)]\quad.\nonumber
\end{equation}
So far, the stochastic covariant result is met, but it can only be regarded as a special case.

\subsubsection{Our objection}
Our objection against Onsager-Machlup Lagrangians, which deviate from our version Eq. (\ref{General8}) is twofold. In order to avoid confusion, we address covariance in curved spaces and stochastic covariance separately. Both fields are not related, and to merge them would only enhance the difficulties of the topic.

\emph{Covariance in curved spaces}: We have not found a reason for why our results are not completely met by covariance-treatments. However, as our derivation is straight-forward, we have best reasons to be convinced that our version is the correct one. Our result establishes a reversible connection between the Fokker-Planck equation and the Onsager-Machlup Lagrangian. By dropping the third term in Eq. (\ref{General8}), this reversible connection would break. If the Fokker-Planck equation is covariant, which it is, then our result automatically meets this condition, too. In conclusion, we have put our horse before the cart and drove the path Fokker-Planck equation $\rightarrow$ Onsager-Machlup Lagrangian. Only Dekker \cite{Dekker1}, by a completely different approach, meets our results.

\emph{Stochastic covariance}: Here we at first must remind ourselves that initially the Fokker-Planck equation and the Langevin-equation have not been related to each other. This relation had been assumed, but it is solely Itô's lemma, which validates this assumption. Thus, on the route Fokker-Planck equation $\rightarrow$ Onsager-Machlup Lagrangian Itô's lemma is an access road, which establishes a connection to stochastic calculus. When Itô's lemma is not invoked the transition-amplitude Eq. (\ref{General7}), the Onsager-Machlup Lagrangian Eq. (\ref{General8}) and the Fokker-Planck equation Eq. (\ref{General2}) describe classical propagation, classical fields and trajectories as any other Lagrangian or differential-equation with physical meaning does. This says that neither the Fokker-Planck equation, nor the Onsager-Machlup Lagrangian need stochastic calculus in order to have a meaning. To our regards, here again the cart is put before the horse, since it is quite the contrary. Stochastic calculus is in need of the Fokker-Planck equation in order to assign a sound, beyond phenomenological meaning to the Langevin-equation in a mathematical sense.

Well understood, we do not say that we doubt about the Langevin-equation to be the equation of motion for stochastic processes. However, Itô's version of stochastic calculus is not unique, and thus also the structure of the Langevin-equation is not unique. Other versions of stochastic calculus are possible, of which the best known is Stratonovich's version. The problem of how to arrange things plain here is solely on the side of mathematics, from the perspective of physics things are sound.

Getting mathematical things plain here must be done by a construction of a covariant stochastic Leibniz-rule, that leaves the transition-amplitude Eq. (\ref{General7}) and the Onsager-Machlup Lagrangian Eq. (\ref{General8}) invariant. If the third term in the Onsager-Machlup Lagrangian Eq. (\ref{General8}) is artificially dropped then, as above, the connection to the Fokker-Planck equation breaks. First attempts to construct a covariant Leibniz-rule on curved spaces have already been undertaken by Cugliandolo et. al. \cite{Cugliandolo2}, but still there is work left to do.
\newline

At their core, all covariance-treatments pursue to derive the Onsager-Machlup Lagrangian from first principles. To our regards, the first principle solely is the Boltzmann-equation, from which a straight-forward route leads to the Onsager-Machlup Lagrangian Eq. (\ref{General8}). First principles based on other fields are irrelevant for this connection. If they lead to different results, then also the Boltzmann-equation as the only first principle of importance is questioned. On the base of this objection, we keep the third term in the Lagrangian Eq. (\ref{General8}) as we have soundly and straight-forwardly derived it.

\subsection{Single-particle Lagrangian}
In order to obtain the Lagrangian, that describes the motion of a single particle in a disordered and thermal environment, we use Ornstein's fluctuation-dissipation theorem, $\sigma_{2}=k_{{\rm{B}}}\,T/m=\beta^{-1}/m$, such that we can write the transition-amplitude by
\begin{equation}
G(q,\,t|q_{0},\,t_{0})\,=\,\int\,\mathcal{D}[q(t)]\,\mathcal{N}[q(t)]\,\exp\left[-\,\beta\,\int_{t_{0}}^{t}d\tau\,\mathcal{L}(\dot{q},\,q)\right]\quad.
\label{General10}\end{equation}
This setting leads to the single-particle Lagrangian
\begin{equation}
\mathcal{L}(\dot{q},\,q)\,=\,\frac{m}{2}\,\frac{\dot{q}(t)^{2}}{b[q(t)]^{2}}\,+\,\beta^{-1}\,\dot{q}(t)\,V_{1}[q(t)]\,-\,\beta^{-1}\,V_{2}[q(t)]\,-\,\frac{\beta^{-2}}{m}\,V_{3}[q(t)]\quad,
\label{General11}\end{equation}
with the gyroscopic potential
\begin{equation}
V_{1}[q(t)]\,=\,2\,\frac{\partial_{q}\,b[q(t)]}{b[q(t)]}-\,m\,\beta\,\frac{a[q(t)]}{b[q(t)]^{2}}\quad,
\label{General12}\end{equation}
a potential, that only contributes for a non-zero drift-field $a[q(t)]$,
\begin{equation}
V_{2}[q(t)]\,=\,2\,a[q(t)]\,\frac{\partial_{q}\,b[q(t)]}{b[q(t)]}\,-\,\beta\,\frac{m}{2}\,\frac{a[q(t)]^{2}}{b[q(t)]^{2}}\,-\,\partial_{q}\,a[q(t)]\quad,
\label{General13}\end{equation}
and a potential solely due to the environment $b[q(t)]$,
\begin{equation}
V_{3}[q(t)]\,=\,b[q(t)]\,\partial_{q}^{2}\,b[q(t)]\,-\,\left(\partial_{q}\,b[q(t)]\right)^{2}\quad.
\label{General14}\end{equation}
So far, we find potentials, which describe the influence of the environment by the functions $a[q(t)],\,b[q(t)]$. Note that in higher dimensions the gyroscopic term is a vector-potential.

\subsection{Inclusion of interaction}
In analogy to classical mechanics, two types of interaction can be introduced. An external interaction $U_{\rm{E}}$  describes forces, which act upon the system from the outside, while an internal interaction $U_{\rm{I}}$ is modulated by $b[q(t),\,t]$, since this function describes properties of the environment the system is immersed in. Thus, the Lagrangian should read
\begin{eqnarray}
\mathcal{L}(\dot{q},\,q)&=&b[q(t),\,t]^{-2}\,\left(\frac{m}{2}\,\dot{q}(t)^{2}\,-\,U_{\rm{I}}[q(t)]\right)\,+\,\beta^{-1}\,\dot{q}(t)\,V_{1}[q(t),\,t]\nonumber\\
&&-\,\beta^{-1}\,V_{2}[q(t),\,t]\,-\,\frac{\beta^{-2}}{m}\,V_{3}[q(t),\,t]\,-\,U_{\rm{E}}[q(t)]\quad.
\label{General15}\end{eqnarray}
Note that we have generalized the functions $a[q(t),\,t],\,b[q(t),\,t]$ by an explicit time-argument, but keep the interactions $U_{\rm{I,\,E}}[q(t)]$ conservative. Our ansatz of how to include interaction is purely motivated by physical reasoning, however, we remind the reader that now the function $b[q(t),\,t]$ in the first term of the Lagrangian acts like a Helmholtz-factor.

The motivation for our construction can be read off from the equation of motion,
\begin{equation}
\ddot{q}(t)\,=\,F_{\rm{pot}}[q(t),\,t]\,+\,F_{\rm{env}}[q(t),\,t]\,+\,F_{\rm{diss}}[\dot{q}(t),\,q(t),\,t]\quad,
\label{General16}\end{equation}
where $F_{\rm{pot}}[q(t),\,t]$ is the force, that stems from the potentials, $F_{\rm{env}}[q(t),\,t]$ is the force, that is generated by the environment and $F_{\rm{diss}}[q(t),\,t]$ is the force, that generates dissipation. The force due to the interaction is given by
\begin{equation}
F_{\rm{pot}}[q(t),\,t]\,=\,-\,\frac{1}{m}\,\left(\frac{\partial\,U_{I}[q(t)]}{\partial\,q(t)}\,-\,2\,U_{I}[q(t)]\,\frac{\partial\ln[b[q(t),\,t]]}{\partial\,q(t)}\,+\,b[q(t),\,t]^{2}\,\frac{\partial\,U_{E}[q(t)]}{\partial\,q(t)}\right)\quad,
\label{General17}\end{equation}
the force due to the environment reads
\begin{equation}
F_{\rm{env}}[q(t),\,t]\,=\,-\,b[q(t),\,t]^{2}\,\left(\frac{\beta^{-1}}{m}\,\frac{\partial\,V_{1}[q(t),\,t]}{\partial\,t}\,+\,\frac{\beta^{-1}}{m}\,\frac{\partial\,V_{2}[q(t),\,t]}{\partial\,q(t)}\,+\,\frac{\beta^{-2}}{m^{2}}\,\frac{\partial\,V_{3}[q(t),\,t]}{\partial\,q(t)}\right)\quad,
\label{General18}\end{equation}
and last, the dissipative force is
\begin{equation}
F_{\rm{diss}}[\dot{q}(t),\,q(t),\,t]\,=\,\,2\,\frac{\partial\ln\left[b[q(t),\,t]\right]}{\partial\,t}\,\dot{q}(t)\,+\,\frac{\partial\ln\left[b[q(t),\,t]\right]}{\partial\,q(t)}\,\dot{q}(t)^{2}\quad.
\label{General19}\end{equation}
Strictly spoken the forces in our notation are accelerations, but we keep it like this for better understanding and more insight into the structure.

By inspection of Eq. (\ref{General17}), we understand that our above setting for the inclusion of potentials should be correct. The classical force generated by the internal potential $U_{\rm{I}}[q(t)]$ is not be modulated by the properties of the environment as described by $b[q(t),\,t]$, but the classical external force is very well be modulated. However, the second term of the force $F_{\rm{pot}}[q(t),\,t]$ clarifies that the environment of course also couples to the internal potential and creates an additional non-classical force. The environmental force $F_{\rm{env}}[q(t),\,t]$, Eq. (\ref{General18}), is solely determined by $a[q(t),\,t]$ and $b[q(t),\,t]$. Furthermore, we see that the gyroscopic term $V_{1}[q(t),\,t]$ only contributes for explicit time-dependence. Finally, the dissipative force $F_{\rm{diss}}[\dot{q}(t),\,q(t),\,t]$, Eq. (\ref{General19}), contains terms that directly couple to the velocity, and we see that Stokes-friction and Newtonian friction are naturally related to the environmental function $b[q(t),\,t]$. This was to be expected, since the structure of the dissipation should depend on the properties of the environment.

From $F_{\rm{diss}}[q(t),\,t]$, Eq. (\ref{General19}), we easily deduce that constant Stokes-friction is generated by an environmental function $b(t)\,=\,\exp[\mu_{1}\,t]$, while constant Newtonian friction follows from $b[q(t)]\,=\,\exp[\mu_{2}\,q(t)]$.

As said in the introduction, we now easily see that the Caldirola-Kanai Hamiltonian \cite{Caldirola, Kanai} belongs to the Onsager-Machlup family with Helmholtz-factor $b(t)=\exp[\mu_{1}/2\,t]$,
\begin{equation}
\mathcal{H}(p,\,q)\,=\,\exp[\mu_{1}\,t]\,\frac{p(t)^{2}}{2\,m}\,+\,\exp[-\,\mu_{1}\,t]\,\frac{m}{2}\,\omega^{2}\,q(t)^{2}\quad.\nonumber
\end{equation}

\subsection{Additive and multiplicative structure of the environment}
The function $b[q(t),\,t]$ describes the properties of the environment, and we now shall make some general statements about it's structure.

The environmental function may be assumed to be additive, $b[q(t),\,t]=\sum_{i}^{n}b_{i}[q(t),\,t]$, if for a certain $b_{i}[q(t),\,t]=1$ or even $b_{i}[q(t),\,t]=0$ the structure of the system changes essentially. A multiplicative coupling, $b[q(t),\,t]=\prod_{i}^{n}b_{i}[q(t),\,t]$, may be given if for a certain  $b_{i}[q(t),\,t]=1$ no essential change in the structure of the system does occur. This can be understood by considering the friction-terms in Eq. (\ref{General19}).

In the case of a multiplicative connection of Stokes friction and Newtonian friction, $b[q(t),\,t]=b_{1}(t)\,b_{2}[q(t)]=\exp[\mu_{1}\,t]\exp[\mu_{2}\,q(t)]$, both contributions of friction are constant, and there is no essential change if one multiplicand is set equal to unity. By inspection of the equation of motion,
\begin{equation}
\ddot{q}(t)\,=\,2\,\mu_{1}\,\dot{q}(t)\,+\,\mu_{2}\,\dot{q}(t)^{2}\,+\,...\quad,\nonumber
\end{equation}
we see that for either $\mu_{1}=0$ or $\mu_{2}=0$ the dynamic changes, but not the structure of the equation of motion.

In the case of an additive connection, e.g. $b[q(t),\,t]=b_{1}(t)+b_{2}[q(t)]=\exp[\mu_{1}\,t]+\exp[\mu_{2}\,q(t)]$ the equation of motion reads
\begin{equation}
\ddot{q}(t)\,=\,\frac{2\,\mu_{1}\,\exp[\mu_{1}\,t]\,\dot{q}(t)\,+\,\mu_{2}\,\exp[\mu_{2}\,q(t)]\,\dot{q}(t)^{2}}{\exp[\mu_{1}\,t]+\exp[\mu_{2}\,q(t)]}\,+\,...\quad.\nonumber
\end{equation}
If now either $\mu_{1}=0$ or $\mu_{2}=0$ the dynamic changes, if but either $b_{1}(t)$ or $b_{2}[q(t)]$ is set equal to zero the system changes essentially, since then also the equation of motion changes essentially.

\section{Motion through a gaseous medium with external field}
In the following, we explore examples of motion through a barometric medium. We set $a[q(t),\,t]=0$ and focus on the function $b[q(t),\,t]$. The dynamics, that is generated by $b[q(t),\,t]$ is rich and opens the possibility to treat problems, which are not easily accessible by other methods. Besides the free fall, we also explore the possibility to model a blow of wind by a solitonic pulse, that runs through the medium.

\subsection{Barometric medium}
An important example for Newtonian friction is the free fall from a certain height $h_{0}$ through a gaseous medium in a gravitational field. The well-known phenomenological equation of motion is given by
\begin{equation}
\ddot{h}(t)\,=\,-\,g\,+\,\mu_{2}\,\exp\left[-\,\beta\,m\,g\,h(t)\right]\,\dot{h}(t)^{2}\quad,
\label{FF1}\end{equation}
where the exponential term is the barometric formula. Note that we use the gravitational acceleration $g>0$\,, which explains the signs in the equations of motion.

We now make contact with the Lagrangian. The Lagrangian, that describes systems like the present one should read
\begin{equation}
\mathcal{L}(\dot{h},\,h)\,=\,\frac{m}{2}\,\frac{\dot{h}(t)^{2}}{b[h(t)]^{2}}\,-\,m\,g\,h(t)\quad.
\label{FF2}\end{equation}
We employ $b[q(t)]$ as a Helmholtz-factor to resolve the friction-term. Furthermore, it is clear that gravity acts on the gaseous medium as an external interaction. The Lagrangian Eq. (\ref{FF2}) yields the equation of motion
\begin{equation}
\ddot{h}(t)\,=\,-\,g\,b[h(t)]^{2}\,+\,\frac{1}{b[h(t)]}\,\frac{\partial\,b[h(t)]}{\partial\,h(t)}\,\dot{h}(t)^{2}\quad.
\label{FF3}\end{equation}
Thus, we may conclude that we can calculate the environmental function by
\begin{equation}
\frac{1}{b[h(t)]}\,\frac{\partial\,b[h(t)]}{\partial\,h(t)}\,=\,\mu_{2}\,\exp\left[-\,\beta\,m\,g\,h(t)\right]\quad,
\label{FF4}\end{equation}
and consequently, this yields
\begin{equation}
b[h(t)]\,=\,\exp\left[-\,\frac{\mu_{2}}{\beta\,m\,g}\,\exp\left[-\,\beta\,m\,g\,h(t)\right]\right]\quad.
\label{FF5}\end{equation}

The equation of motion, that follows straight forwardly from the Lagrangian Eq. (\ref{FF2}) is then given by
\begin{equation}
\ddot{h}(t)\,=\,-\,g\,\exp\left[-\,2\,\mu_{2}\,T\,\exp\left[-\,h(t)/T\right]\right]\,+\,\mu_{2}\,\exp\left[-\,h(t)/T\right]\,\dot{h}(t)^{2}\quad,
\label{FF6}\end{equation}
where $T^{-1}=\beta\,m\,g$. For $\mu_{2}\,T<1$ we have $b[h(t)]\,\approx\,1$, such that Eq. (\ref{FF1}) is recovered.

We now also include the potential $V_{3}[h(t)]$,
\begin{equation}
\frac{\beta^{-2}}{m}\,V_{3}[h(t)]\,=\,-\,\mu_{2}\,T\,m\,g^{2}\,\exp\left[-\,2\,\mu_{2}\,T\,\exp\left[-\,h(t)/T\right]\,-\,h(t)/T\right]\quad.
\label{FF7}\end{equation}
The criterion for the negligence here is given by $\left(\mu_{2}\,T\right)\,m\,g^{2}<1$\,. The equation of motion follows by
\begin{eqnarray}
\ddot{h}(t)&=&-\,g\,\exp\left[-\,2\,\mu_{2}\,T\,\exp\left[-\,h(t)/T\right]\right]\,+\,\mu_{2}\,\exp\left[-\,h(t)/T\right]\,\dot{h}(t)^{2}\nonumber\\
&&-\,\mu_{2}\,g^{2}\,\exp\left[-\,4\,\mu_{2}\,T\,\exp\left[-\,h(t)/T\right]\,-\,h(t)/T\right]\nonumber\\
&&+\,2\,\mu_{2}^{2}\,T\,g^{2}\,\exp\left[-\,4\,\mu_{2}\,T\,\exp\left[-\,h(t)/T\right]\,-\,2\,h(t)/T\right]\quad.
\label{FF8}\end{eqnarray}

\begin{figure}[t!]\centering\vspace{0.cm}
\rotatebox{0.0}{\scalebox{0.72}{\includegraphics{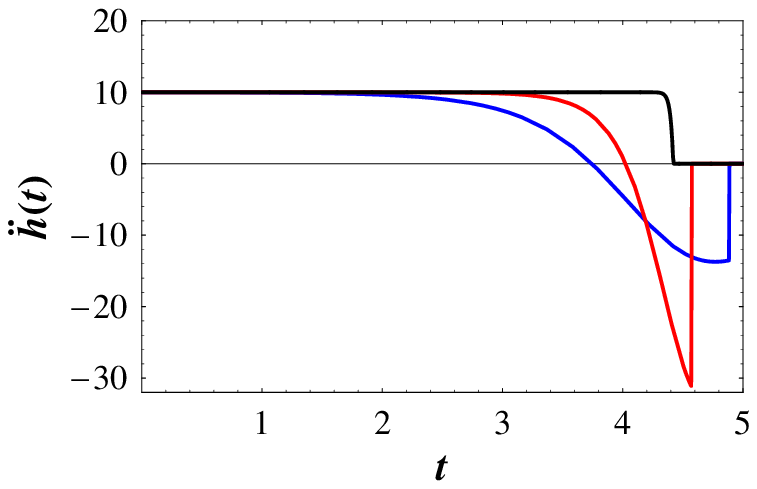}}}
\rotatebox{0.0}{\scalebox{0.72}{\includegraphics{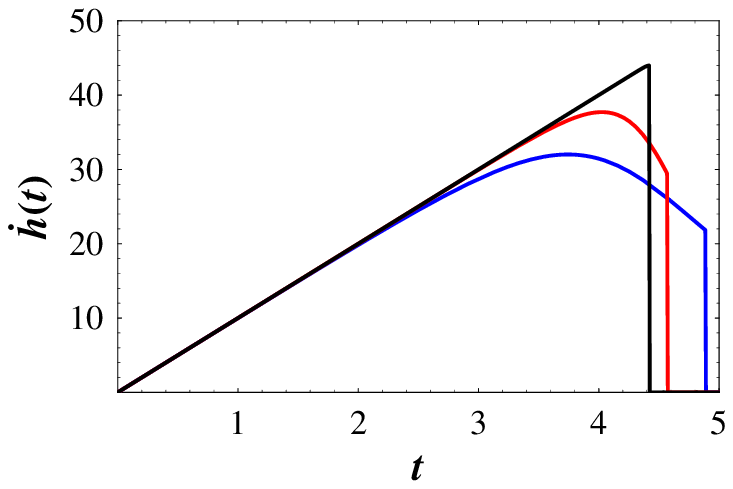}}}
\rotatebox{0.0}{\scalebox{0.72}{\includegraphics{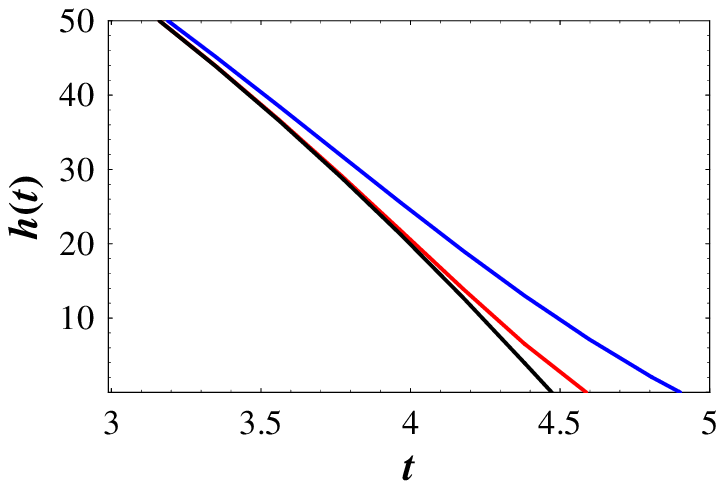}}}
\rotatebox{0.0}{\scalebox{0.72}{\includegraphics{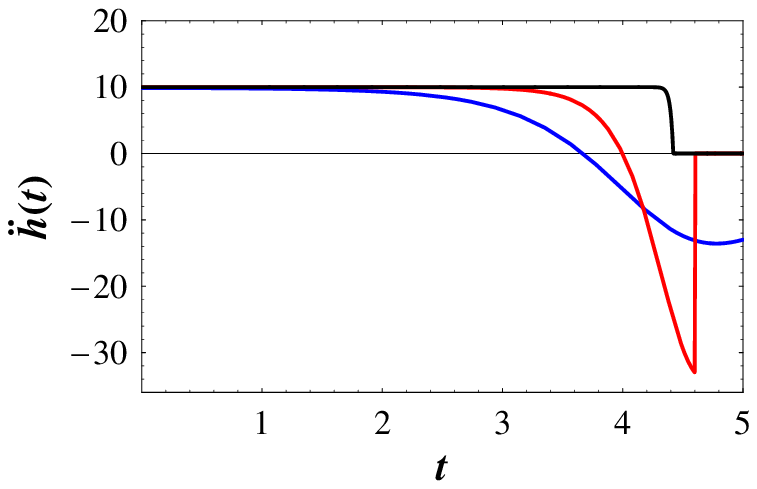}}}
\rotatebox{0.0}{\scalebox{0.72}{\includegraphics{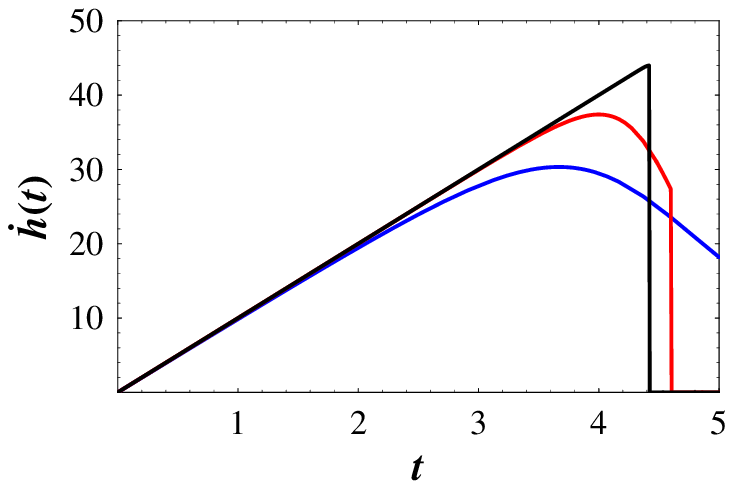}}}
\rotatebox{0.0}{\scalebox{0.72}{\includegraphics{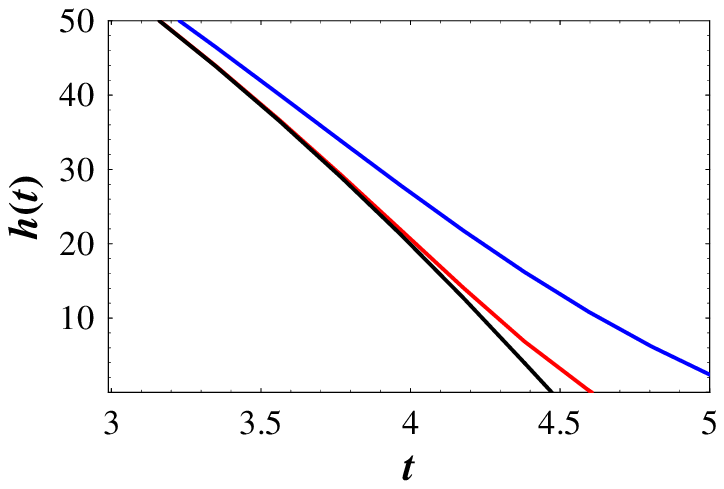}}}
\rotatebox{0.0}{\scalebox{0.72}{\includegraphics{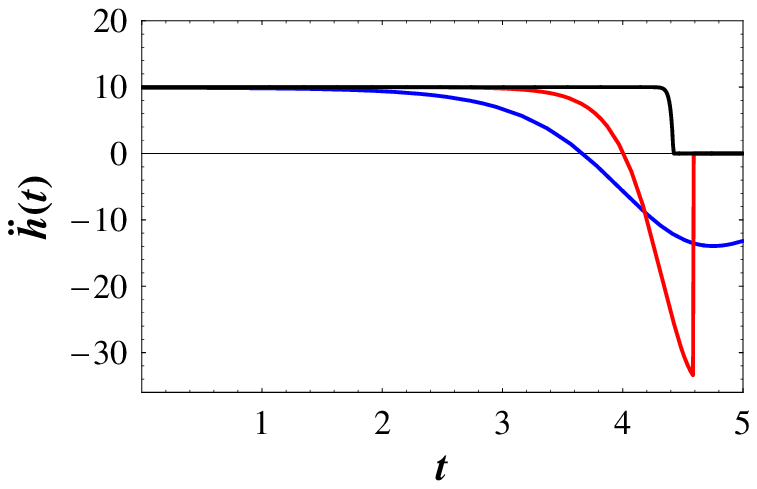}}}
\rotatebox{0.0}{\scalebox{0.72}{\includegraphics{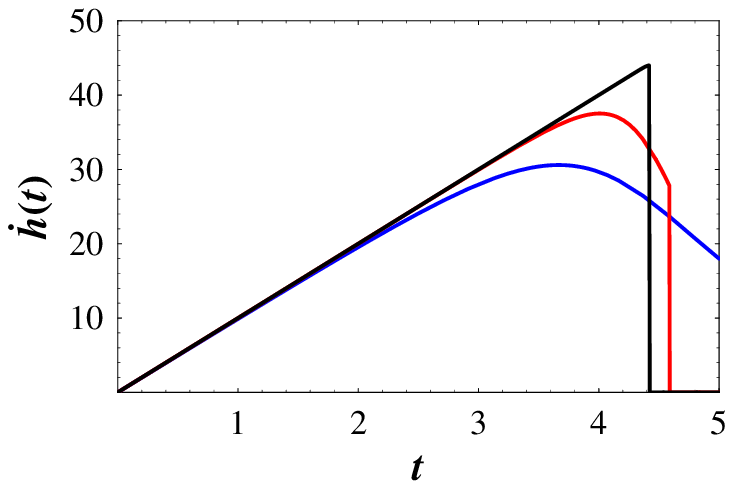}}}
\rotatebox{0.0}{\scalebox{0.72}{\includegraphics{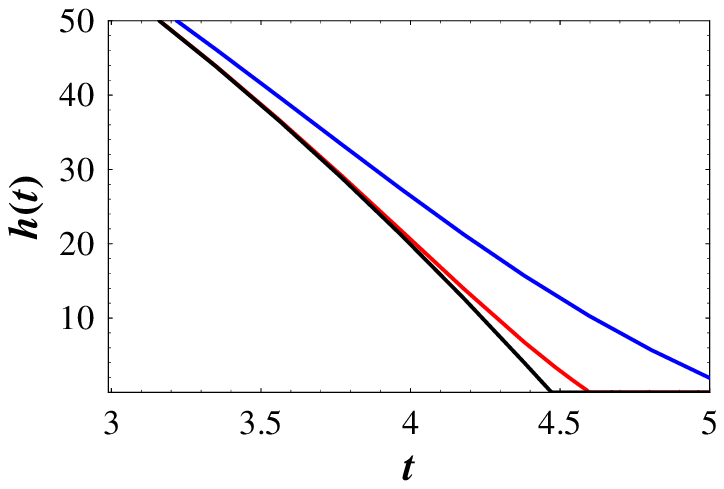}}}
\caption{\footnotesize{Free fall according to Eq. (\ref{FF1}) (upper row), Eq. (\ref{FF6}) (middle row) and Eq. (\ref{FF8}) (lower row), initial velocity $\dot{h}_{0}=0$, initial height $h_{0}=100$, $g=10$, $\mu_{2}=0.05$, colours: $T=\{$1 (black),\,10 (red),\,20 (blue)$\}$.}}
\label{fig21}\end{figure}

In Fig. (\ref{fig21}) we observe that the dynamics for $T=\{1,\,10\}$ (black and red curves) is equal for all three equations of motion Eqs. (\ref{FF1},\,\ref{FF6},\,\ref{FF8}). For $T=20$, thus $\mu_{2}\,T=1$ (blue curves), we but find that the dissipation described by Eqs. (\ref{FF6}, \ref{FF8}) is considerably enhanced. Since it holds $\mu_{2}\,T=1$, we conclude that the phenomenological equation of motion Eq. (\ref{FF1}) is not valid in this regime, and one has to work with the full equation of motion Eq. (\ref{FF8}) instead. By definition, $T^{-1}=\beta\,m\,g$, the deviation should be related to mass and temperature. Consequently, we find that the resistivity of a barometric medium against directional motion for higher temperatures and smaller masses is larger than expected by the phenomenological equation of motion Eq. (\ref{FF1}). We are unaware if this effect is confirmed by e.g. hydrodynamic approaches. However, if this effect is not to be found, then it is likely that gravity acts like an internal interaction. To us this would be counter-intuitive.

\subsection{Free fall with a blow of wind}
We now treat the free fall through a homogeneous medium, $b_{0}[q(t)]\,=\,1$, distorted by a localized pulse, that may be interpreted as a blow of wind. The question is how this system can be modeled. The first thought may certainly be that an additional potential should be added, however, just from geometric reasoning we understand that this strategy must fail, since we cannot locate this potential on the way down. The remaining strategy is to treat the pulse as an additional property of the medium itself. A blow of wind runs through the medium as a longitudinal density-wave, which we chose to model by a solitonic pulse. Hence, the properties of the medium should be described by
\begin{equation}
b[h(t),\,t]\,=\,b_{0}[h(t)]\,+\,b_{1}[h(t),\,t]\,=\,1\,+\,\frac{A}{\cosh\left[k\,(h(t)\,-\,h_{1})\,-\,\omega\,t\right]^{2}}\quad.
\label{FFW1}\end{equation}
The sign of $A$ decides about the direction of the pulse, and $h_{1}$ is the height where the center of pulse starts to propagate. Our system provides an example for an additive connection of the function $b[q(t),\,t]$. If the function $b[q(t),\,t]$ would be multiplicative, the system would have exotic properties, since gravity would only take effect while the particle passes through the center of the pulse.

To be more realistic, we furthermore assume that the medium the motion takes place in is barometric. The environmental function thus reads
\begin{equation}
b[h(t),\,t]\,=\,\exp\left[-\,\mu_{2}\,T\,\exp\left[-\,h(t)/T\right]\right]\left(1\,+\,\frac{A}{\cosh\left[k\,(h(t)\,-\,h_{1})\,-\,\omega\,t\right]^{2}}\right)\quad.
\label{FFW2}\end{equation}
Here now a global multiplicative connection applies, because the pulse of course must couple to the barometric properties. If we set $b_{1}[h(t),\,t]=0$ the effect of the pulse vanishes, and we are left with an ordinary barometric medium. This illustrates an essential change of the properties of the system as discussed above.

Since the environmental function $b[h(t),\,t]$ is explicitly time-dependent, the gyroscopic term of the potential contributes, see Eq. (\ref{General11}). Hence, the Lagrangian is given by
\begin{equation}
\mathcal{L}(\dot{h},\,h)\,=\,\frac{m}{2}\,\frac{\dot{h}(t)^{2}}{b[h(t),\,t]^{2}}\,+\,2\,\dot{h}(t)\,m\,g\,T\,V_{1}[h(t),\,t]\,-\,m\,g^{2}\,T^{2}\,V_{3}[h(t),\,t]-\,m\,g\,h(t)\quad.
\label{FFW3}\end{equation}
The equation of motion is lengthy and cumbersome, and we shall not write it down.

\begin{figure}[t!]\centering\vspace{0.cm}
\rotatebox{0.0}{\scalebox{0.88}{\includegraphics{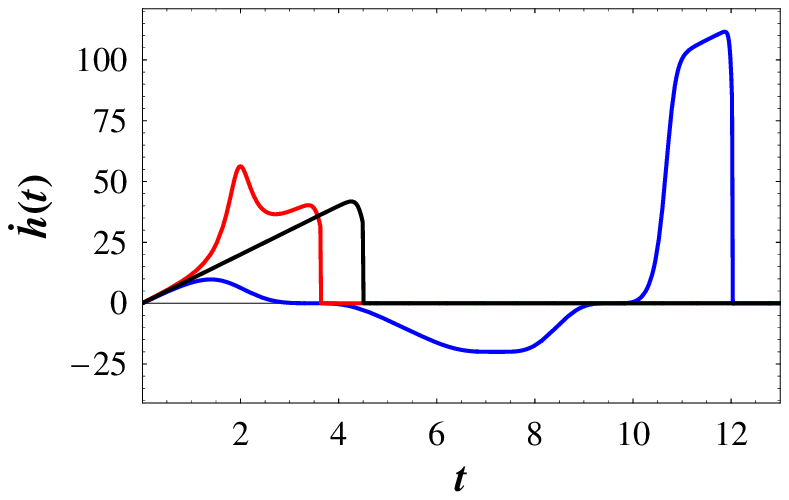}}}
\rotatebox{0.0}{\scalebox{0.85}{\includegraphics{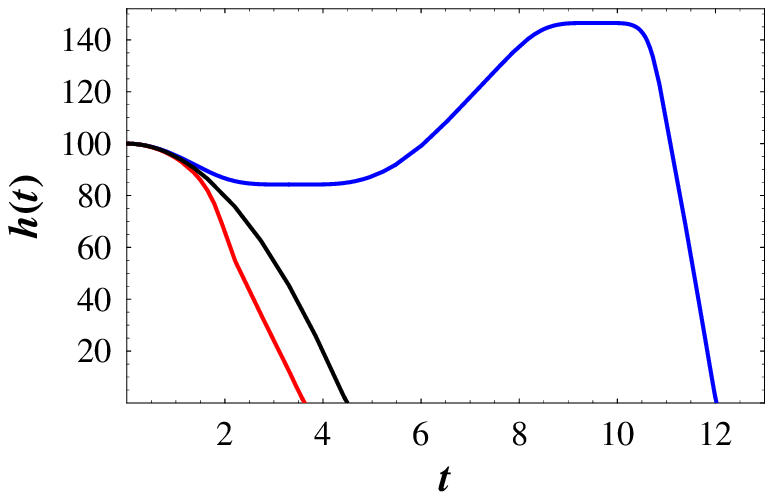}}}
\caption{\footnotesize{Free fall with a blow of wind in a barometric medium. Initial velocity $\dot{h}_{0}=0$, initial pulse-height $h_{1}=50$, initial height $h_{0}=100$, $g=10$, wave-vector $k=0.05$, frequency $\omega=0.5$, $T=3.333$, $\mu_{2}=0.097$. Colours: $A=\{$0 (black),\,1.0 (red),\,$-$\,1.0 (blue)$\}$.}}
\label{fig23}\end{figure}

In Fig. (\ref{fig23}), we illustrate the effects of the environmental function $b[h(t),\,t]$ according to Eq. (\ref{FFW2}). As long as the falling object is distorted by the solitonic pulse, the motion considerably deviates from it's normal behaviour. A blow of wind downwards accelerates the fall beyond the gravitational acceleration, while a blow of wind from below depletes the acceleration and delays the motion considerably. Consequently, we may conclude that our assumption about the  of the function $b[q(t),\,t]$ is correct, and that a solitonic pulse in principal is a reasonable model for a blow of wind. We remark that the effect of a pulse from above does not so sensitively depend on the values of the parameters as a pulse from below. Already a small change in the parameters can change the behaviour of the trajectory. For some values, the retardation is even larger, for others the retardation is smaller. This effect clearly is due to the high non-linearity of the equation of motion, and also due to the model for the pulse. However, our present example demonstrates that dynamics based on the Onsager-Machlup Lagrangian can describe scenarios and systems, which else are not accessible by classical mechanics.

\section{Harmonic oscillator}
For the harmonic oscillator we determine the environmental function $b[q(t)]$ by the same approach as we have done for the barometric medium above. Assuming that the friction-term in a bath of oscillators is given by
\begin{equation}
\frac{1}{b[q(t)]}\,\frac{\partial\,b[q(t)]}{\partial\,q(t)}\,=\,\mu_{2}\,\exp\left[-\,q(t)^2/T\right]\,,\quad \,T^{-1}\,=\,\beta\,\frac{m}{2}\,\omega^{2}\quad,
\label{FFO1}\end{equation}
we obtain
\begin{equation}
b[q(t)]\,=\,\exp\left[\frac{\mu_{2}}{2}\,\sqrt{\pi\,T}\,{\rm{erf}}\left[q(t)/\sqrt{T}\right]\right]\quad.
\label{FFO2}\end{equation}
Contrary to the case of gravity above, we chose to work with an internal harmonic interaction here, and the Lagrangian follows by
\begin{eqnarray}
\mathcal{L}(\dot{q},\,q)&=&\left(\frac{m}{2}\,\dot{q}(t)^{2}\,-\,U_{\rm{I}}[q(t)]\right)\,b[q(t)]^{-2}\,-\,\frac{m}{4}\,\omega^{4}\,T^{2}\,V_{3}[q(t)]\nonumber\\
&=&\frac{m}{2}\,\left(\dot{q}(t)^{2}\,-\,\omega^{2}\,q(t)^{2}\right)\,\exp\left[-\,\mu_{2}\,\sqrt{\pi\,T}\,{\rm{erf}}\left[q(t)\,/\sqrt{T}\right]\right]\nonumber\\
&&+\,\frac{m}{2}\,\omega^{4}\mu_{2}\,T\,q(t)\,\exp\left[\mu_{2}\,\sqrt{\pi\,T}\,{\rm{erf}}\left[q(t)/\sqrt{T}\right]\,-\,q(t)^{2}/T\right]\quad.
\label{FFO3}\end{eqnarray}
Note that so far the gyroscopic term can be neglected, because the function $b[q(t)]$ is not explicitly time-dependent. In analogy to our results above, we find that the thermal corrections are small for $\left(\mu_{2}\,T\right)\,m\,\omega^{4}<1$\,.

\begin{figure}[t!]\centering\vspace{0.cm}
\rotatebox{0.0}{\scalebox{0.99}{\includegraphics{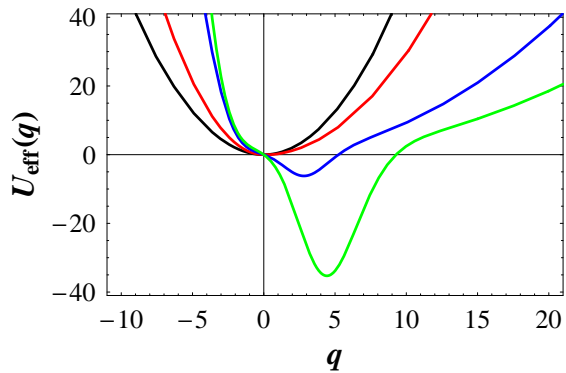}}}
\caption{\footnotesize{Effective potential $U_{\rm{eff}}[q(t)]$. Mass $m=1$, frequency $\omega=1$, friction $\mu_{2}=0.3$, unpertubed oscillator (black curve). Red curve: $T=1$, blue curve: $T=10$, green curve: $T=20$.}}
\label{fig31}\end{figure}

In Fig. (\ref{fig31}) we illustrate the effective potential,
\begin{equation}
U_{\rm{eff}}[q(t)]\,=\,U_{\rm{I}}[q(t)]\,b[q(t)]^{-2}\,+\,\frac{m}{4}\,\omega^{4}\,T^{2}\,V_{3}[q(t)]\quad,
\label{FFO4}\end{equation}
for several values of $T$, in comparison with the unpertubed oscillator $U_{\rm{I}}[q(t)]$. We see that for $\{\mu_{2}>0,\, T>0\}$ the effective potential becomes skew on the right side. For $\mu_{2}<0$ but the skewness is left-sided. This allows us to interpret the friction $\mu_{2}$ also in the sense of an order-parameter, since it's sign decides on which side the system will condense. Last, the higher the thermal energy $T$, the deeper lies the minimum of the well.

\subsection{Inert thermal environment}
An inert thermal environment is given if the dissipation of energy does not enhance the thermal energy $T$ of the environment. This implies that the environment is large enough to absorb dissipated energy without back-reaction.

\begin{figure}[t!]\centering\vspace{0.cm}
\rotatebox{0.0}{\scalebox{0.72}{\includegraphics{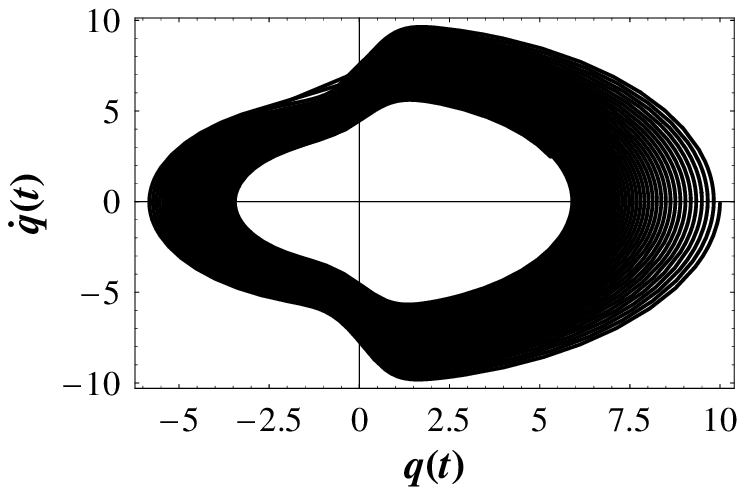}}}
\rotatebox{0.0}{\scalebox{0.72}{\includegraphics{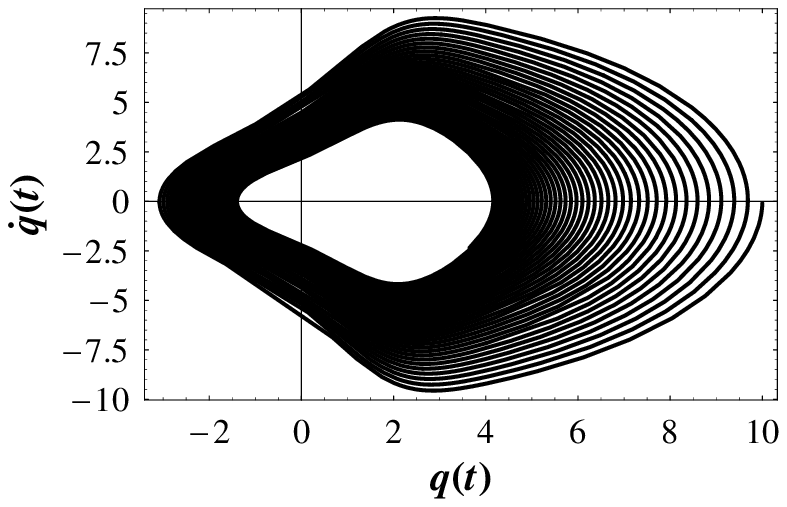}}}
\rotatebox{0.0}{\scalebox{0.72}{\includegraphics{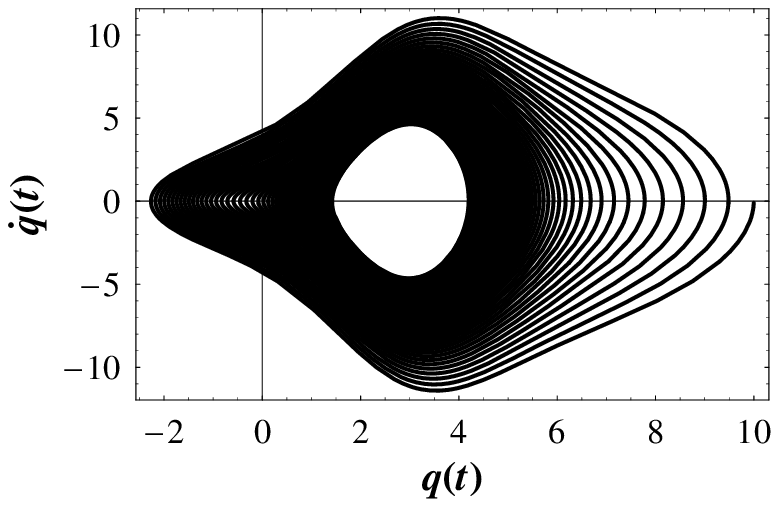}}}
\caption{\footnotesize{Phase-space of the damped oscillator calculated for 250 time-units. Parameters: $\{m=1, \omega=1, \mu_{2}=0.3\}$. Initial conditions are $q_{0}=10$, $\dot{q}_{0}=0$. From left to right: $\{T=1, T=5, T=10\}$.}}
\label{fig32}\end{figure}

In Fig. (\ref{fig32}) we show the phase-space $(\dot{q},\,q)$ of the damped oscillator with friction $\mu_{2}=0.3$ for three thermal energies $\{T=1,\,T=5,\,T=10\}$. Despite the friction $\mu_{2}$ has the same value for all three cases, we deduce that the thermal energy considerably influences the behaviour of the trajectories and velocities. The skew shape of the effective potential is clearly visible, and furthermore, we see that the deeper the well of the effective potential is, the faster the particle drops into the well. The behaviour found matches with the behaviour of a particle, that moves through a barometric medium, see above. If we identify the thermal energy with noise or disorder, we easily understand that the more disorder is present, the more the motion becomes impeded. Remind that the trajectory $q(t)$ is the most probable path a particle takes through the environment.

The energy as a function of time, $E(t)$, to analyze the dissipation is not to be confused with the kinetic Hamiltonian Eq. (\ref{General6}). The total energy we need here stems from what we call the dynamic Hamiltonian, that follows from the Lagrangian Eq. (\ref{FFO3}) by the usual Legendre-transform. The dynamic Hamiltonian thus corresponds to the ordinary total energy
\begin{equation}
E(t)\,=\,\frac{m}{2}\,\frac{\dot{q}(t)^{2}}{b[q(t)]^{2}}\,+\,U_{\rm{eff}}[q(t)]\quad.
\label{FFO7}\end{equation}

\begin{figure}[t!]\centering\vspace{0.cm}
\rotatebox{0.0}{\scalebox{0.99}{\includegraphics{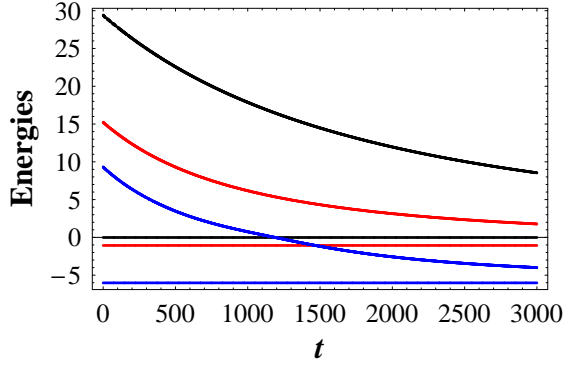}}}
\caption{\footnotesize{Time-dependent energies of the damped oscillator. Parameters: $\{m=1,\, \omega=1,\, \mu_{2}=0.3\}$. Initial conditions are $q_{0}=10$,\, $\dot{q}_{0}=0$\,. Colours: $T=1$ (black), $T=5$ (red), $T=10$ (blue). The horizontal lines mark the minimum of the effective potential $U_{\rm{eff}}(q_{\rm{min}})$, respectively.}}
\label{fig33}\end{figure}

From Fig. (\ref{fig33}) we deduce that the dissipation of energy is a monotonous process, and the functions $\{E(t)\}$ approach the minimum of the effective potential $U_{\rm{eff}}(q_{\rm{min}})$ asymptotically. This matches with the phase-space depicted in Fig. (\ref{fig32}). So far, a harmonic oscillator with friction in a disordered environment behaves exactly as we would expect it. 

\subsection{Inclusion of back-reaction and dynamical balance}
The back-reaction of the environment can be included by allowing the thermal energy $T\,\rightarrow\,T(t)$ to be time-dependent. By doing so, the gyroscopic term has to be taken into account. The Lagrangian thus reads
\begin{equation}
\mathcal{L}(\dot{q},\,q)\,=\,\frac{m}{2}\,\frac{\dot{q}(t)^{2}}{b[q(t),\,T(t)]^{2}}\,-\,\frac{U_{\rm{I}}[q(t)]}{b[q(t),\,T(t)]^{2}}+\,\dot{q}(t)\,m\,\omega^{2}\,T(t)\,V_{1}[q(t),\,T(t)]\,-\,\frac{m}{4}\,\omega^{4}\,T(t)^{2}\,V_{3}[q(t),\,T(t)]\quad.
\label{FFO8}\end{equation}

The force can be divided into four parts. First, we have the force due to the interaction
\begin{equation}
F_{\rm{pot}}[q(t),\,T(t)]\,=\,-\,\frac{1}{m}\,\left(\frac{\partial\,U_{\rm{I}}[q(t)]}{\partial\,q(t)}\,-\,2\,U_{\rm{I}}[q(t)]\,\frac{\partial\,\ln\left[b[q(t),\,T(t)]\right]}{\partial\,q(t)}\right)\quad,
\label{FFO9}\end{equation}
second, we have the force due to the environment,
\begin{equation}
F_{\rm{env}}[q(t),\,T(t)]\,=\,-\,\frac{\omega^{4}}{4}\,T(t)^{2}\,b[q(t),\,T(t)]^{2}\,\frac{\partial\,V_{3}[q(t),\,T(t)]}{\partial\,q(t)}\quad,
\label{FFO10}\end{equation}
third, we have the dissipative force,
\begin{equation}
F_{\rm{diss}}[q(t),\,T(t)]\,=\,\frac{\partial\,\ln\left[b[q(t),\,T(t)]\right]}{\partial\,q(t)}\,\dot{q}(t)^{2}\quad,
\label{FFO11}\end{equation}
and last, we find a thermal force, which of course can also be understood as a force due to the environment,
\begin{eqnarray}
F_{\rm{th}}[q(t),\,T(t)]&=&-\,\omega^{2}\,b[q(t),\,T(t)]^{2}\,\left(T(t)\,\frac{\partial\,V_{1}[q(t),\,T(t)]}{\partial\,T(t)}\,+\,V_{1}[q(t),\,T(t)]\right)\,\dot{T}(t)\nonumber\\
&&+\,2\,\frac{\partial\,\ln\left[b[q(t),\,T(t)]\right]}{\partial\,T(t)}\,\dot{q}(t)\,\dot{T}(t)\quad.
\label{FFO12}\end{eqnarray}
The forces $\{F_{\rm{pot}},\,F_{\rm{env}},\,F_{\rm{diss}}\}$ have the same structure as we have already derived it above, see Eq. (\ref{General13}). The thermal force $F_{\rm{th}}$ contains all terms, that vanish if $T(t)\,=\,{\rm{const.}}$\,, thus $\dot{T}(t)\,=\,0$\,.

The presence of the thermal force opens three possibilities. Either, we can define a function $T(t)$, that describes a heating- or cooling-process controlled from the outside, or,  second, we can treat the system in a self-consistent way. Third, we can do both. Here, we restrict ourselves to a self-consistent treatment.

For a self-consistent treatment, we employ that for each time-step it holds $\Delta E_{i}\,=\,-\,\Delta T_{i}$, since dissipation is present. Thus, the energy $E$ of the system grows if the thermal energy $T$ declines, and vice-verse.

The equation of motion is highly non-linear and the numerics tends to runaway-solutions if the accuracy is not high enough. For the scenarios we present here we have used a time-step of $\Delta t=0.001$. For other choices of the parameters even smaller time-steps are necessary in order to obtain reasonable solutions. This effect is due to the self-consistency of $\Delta T$. If the numerical accuracy is too low $\Delta T$ eventually becomes too large, which then creates runaways. However, some runaways can also be interpreted as a sudden evaporation of the system.

\begin{figure}[t!]\centering\vspace{0.cm}
\rotatebox{0.0}{\scalebox{0.85}{\includegraphics{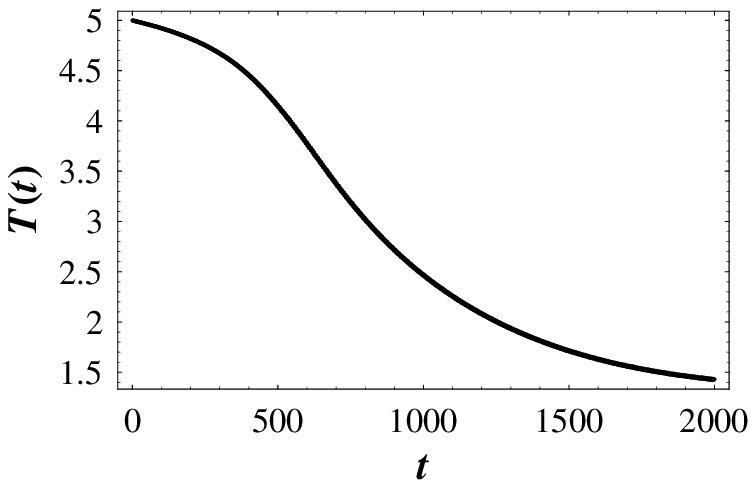}}}
\rotatebox{0.0}{\scalebox{0.85}{\includegraphics{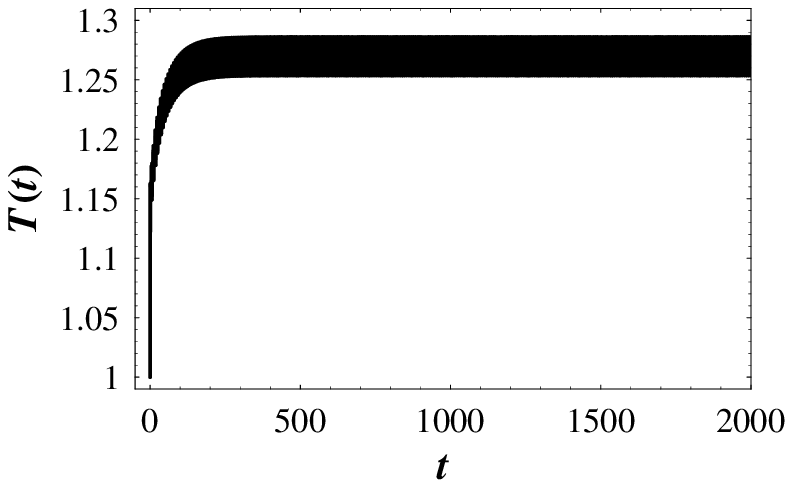}}}
\caption{\footnotesize{Temperatures of the back-reacting system. Parameters: $\{m=1,\, \omega=1,\, \mu_{2}=0.3\}$. Left: initial conditions $q_{0}=1$,\, $\dot{q}_{0}=0$,\, $T_{0}=5$,\, $E_{0}=-0.79$. Right: initial conditions $q_{0}=3.7,$\, $\dot{q}_{0}=0,$ $T_{0}=1$,\, $E_{0}=4.02$.}}
\label{fig34}\end{figure}

In Fig. (\ref{fig34}) we illustrate two examples of different behaviour. The system on the left starts with a thermal energy $T_{0}=5$ and an oscillator-energy of $E_{0}=-0.79$. The time-dependent thermal energy $T(t)$ declines and flows into the oscillator, as it was to be expected. Despite the initial energy-gap is high, $\Delta E=5.79$, the flux of thermal energy into the oscillator takes place only slowly, and after 2000 time-units has still not fully converged towards what we call a dynamical balance. Invisible in Fig. (\ref{fig34}), the function $T(t)$ shows tiny but extremely rapid oscillations, that cannot be interpreted as if the system would converge towards an equilibrium in a strict sense, since they do not decay with time. In many-body-systems close to the thermodynamic limit, the dynamical balance may be interpreted as fluctuations around a real equilibrium.

The system on the right shows a completely different behaviour. Here, the initial energy of the oscillator $E_{0}=4.02$ is higher than the thermal energy $T_{0}=1$. The major influx of energy from the oscillator towards the environment takes place almost instantaneously, and converges fast towards a state of dynamical balance. Again, and this time clearly visible, the thermal energy $T(t)$ shows small but extremely rapid oscillations, that do not decay, but remain steady over time.

We note that for $T_{0}>E_{0}$ the flux of thermal energy towards the oscillator mainly tends to be slow, while in the opposite case, $T_{0}<E_{0}$, the initial exchange of energy mainly tends to take place almost instantaneously. Especially $(T_{0}<E_{0})$-scenarios tend to numerical instabilities. The different behaviour may be explained by the influence of friction. In the case of $T_{0}>E_{0}$ the dissipation of energy from the oscillator towards the environment due to friction acts like a resistance against the influx of energy from the environment towards the oscillator. In the opposite case, $T_{0}<E_{0}$, the initial flux of energy from the oscillator towards the environment increases the action of friction. If $E_{0}$ is too large, which of course depends on the other parameters of the system, the initial flux of energy from the oscillator towards the environment creates an instantaneous jump of $T(t)$, that cannot be recaptured by dynamical balance. In this case the sudden growth of $T(t)$ makes the effective potential $U_{\rm{eff}}[q(t)]$, see Fig. (\ref{fig31}), too attractive, which soon leads to a complete dissipation of the energy of the oscillator.

Furthermore, we note that dynamical balance does not require that oscillator and environment do have the same value of energy, albeit it holds $\Delta T(t)=-\Delta E(t)$, of course. The different values of energy can also be explained by friction. Dynamical balance mainly means that the influence of friction is eliminated. Friction dissipates energy from the oscillator towards the environment, and the same amount of energy flows back from the environment towards the oscillator if, and only if the system is in dynamical balance. The energies $\{T(t),\,E(t)\}$ settle down on values, that buffer friction, enable dynamical balance and thus a stable phase-space. Consequently, the oscillator appears to move frictionless like an ideal oscillator. In conclusion, a small and reactive thermal environment can eliminate the effects of friction and stabilize a system in a state of dynamical balance. Contrary to this, a large and inert thermal environment supports the action of friction. The internal energy of the system $U=E(t)+T(t)=E(0)+T(0)$ of course is conserved.

\section{Two interacting oscillators}
A system of two interacting, harmonically bound particles allows us to study the interaction between two coupled environments. The dynamics of this system is rich, and we shall examine four cases, which we found to be the most interesting ones.

In the first case the harmonic potential is purely external. Oscillator (I) moves in an ideal, frictionless environment, the kinetic energy of oscillator (II) is immersed in a thermal reservoir.

For comparison, the second case is only slightly different from the first case. Here also the harmonic potential of oscillator (II) is internal, and we will see that already this slight change generates dynamics, that is essentially different than in the first case.

In the third case, the setup of the system is the same as in the second case, but we allow a one-sided back-reaction between oscillator (II) and it's environment. For a coupling $\lambda=0.05$ we will discover a behaviour, which may be interpreted as a phase-transition.

Last, in the fourth case both harmonic potentials are internal, and we study the case of two-sided back-reaction between the oscillators and their reservoirs.

In all four cases the harmonic coupling is not immersed into a thermal environment. The harmonic coupling is a mechanical coupling, which we understand to neither belong to environment (I), nor to environment (II). This arrangement may be understood as a mechanical bridge between the two environments.

\subsection{Case I: external harmonic potential}
The Lagrangian in the first case is given by
\begin{eqnarray}
\mathcal{L}\left(\{\dot{q}_{i}\},\,\{q_{i}\}\right)&=&\frac{m_{1}}{2}\,\dot{q}_{1}(t)^{2}\,-\,\frac{m_{1}}{2}\,\omega_{1}^{2}\,\left(q_{1}(t)\,-\,q_{1}\right)^{2}\nonumber\\
&&+\,\frac{m_{2}}{2}\,\frac{\dot{q}_{2}(t)^{2}}{b_{2}\left[q_{2}(t)\,-\,q_{2}\right]^{2}}\,-\,\frac{m_{2}}{4}\,\omega_{2}^{4}\,T_{2}^{2}\,V_{3}^{(2)}[q_{2}(t)\,-\,q_{2}]\,-\,\frac{m_{2}}{2}\,\omega_{2}^{2}\,\left(q_{2}(t)\,-\,q_{2}\right)^{2}\nonumber\\
&&-\,\frac{\lambda^{2}}{2}\,\left(q_{1}(t)\,-\,q_{2}(t)\right)^{2}\quad.
\label{TIP4}\end{eqnarray}
The definition of $T_{2}$ is the same as above, Eq. (\ref{FFO1}).

\begin{figure}[t!]\centering\vspace{0.cm}
\rotatebox{0.0}{\scalebox{0.72}{\includegraphics{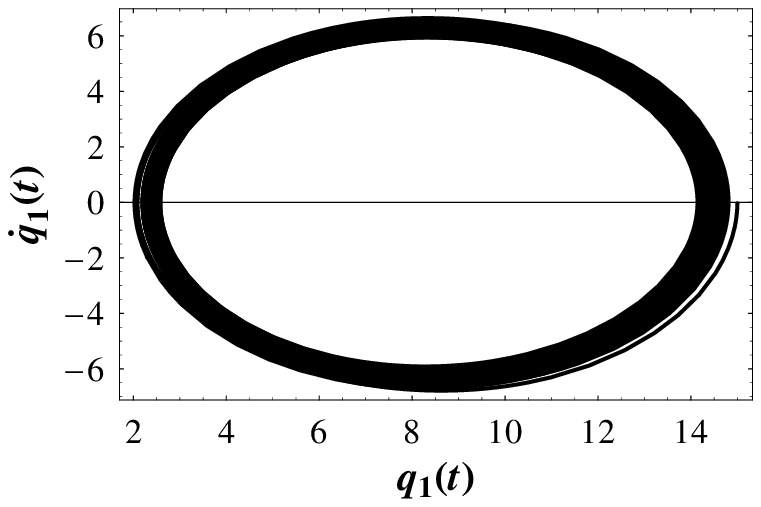}}}
\rotatebox{0.0}{\scalebox{0.72}{\includegraphics{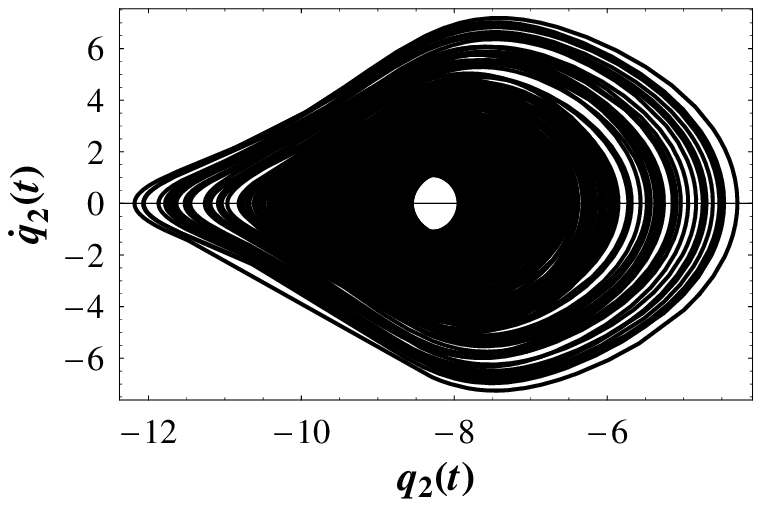}}}
\rotatebox{0.0}{\scalebox{0.72}{\includegraphics{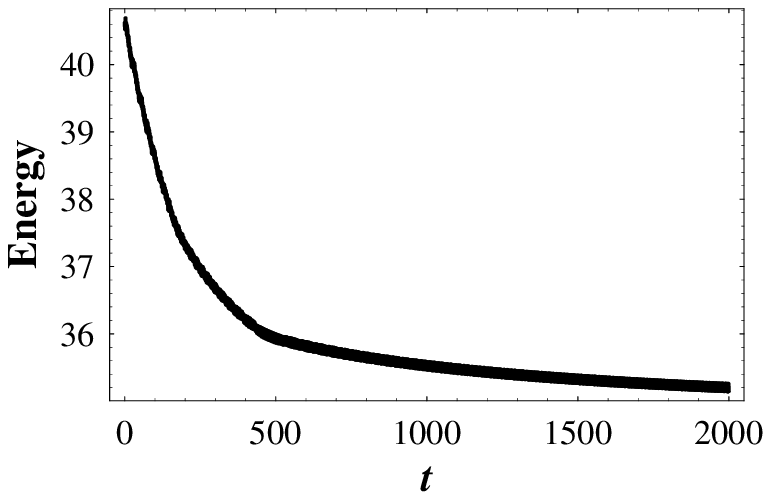}}}
\caption{\footnotesize{Case (I). Phase-space and energy for the inert environment. Oscillator (I) left, oscillator (II) middle. Parameters: $q_{1}=10,\,q_{2}=-10,\,m_{1}=m_{2}=1,\,\omega_{1}=\omega_{2}=1,\, T_{2}=5,\,\mu_{22}=0.3,\,\lambda=0.3$. Initial conditions: $q_{1}(0)=15,\,q_{2}(0)=-10,\, \dot{q}_{1}(0)=\dot{q}_{2}(0)=0$.}}
\label{fig51}\end{figure}

The particle in oscillator (I) moves without friction $\mu_{21}=0$, the particle in oscillator (II) is in contact with a dissipative environment $\{T_{2}=5,\,\mu_{22}=0.3\}$. The harmonic coupling is of strength $\lambda=0.3$\,.

This scenario describes a draining-mechanism. The initial single-particle energy of oscillator (II) is chosen by $E_{2}=0$ in order to have a better illustration of the dissipation of energy.

By Fig. (\ref{fig51}) we see that the phase-space of oscillator (I) very slowly contracts over time, as it was to be expected. After the time-evolution has started, the phase-space of oscillator (II) expands, which reflects an influx of energy from oscillator (I). However, the larger influx but also tends to a faster dissipation, which can be deduced from the time-dependent energy depicted in Fig. (\ref{fig51}) on the right. The dissipation slows down when the phase-space of oscillator (II) starts to contract and ever so slowly approaches the minimum of it's potential. This process is asymptotic, which also can be seen by the behaviour of the energy.

\subsection{Case II: internal harmonic potential}
\begin{figure}[t!]\centering\vspace{0.cm}
\rotatebox{0.0}{\scalebox{0.72}{\includegraphics{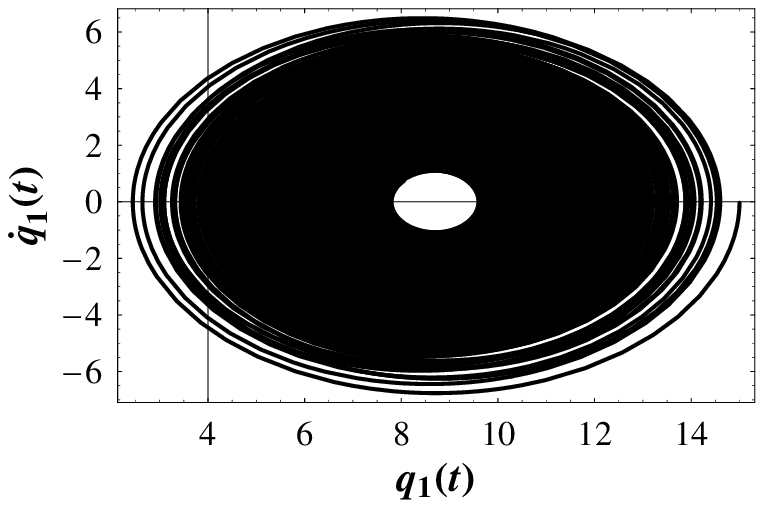}}}
\rotatebox{0.0}{\scalebox{0.72}{\includegraphics{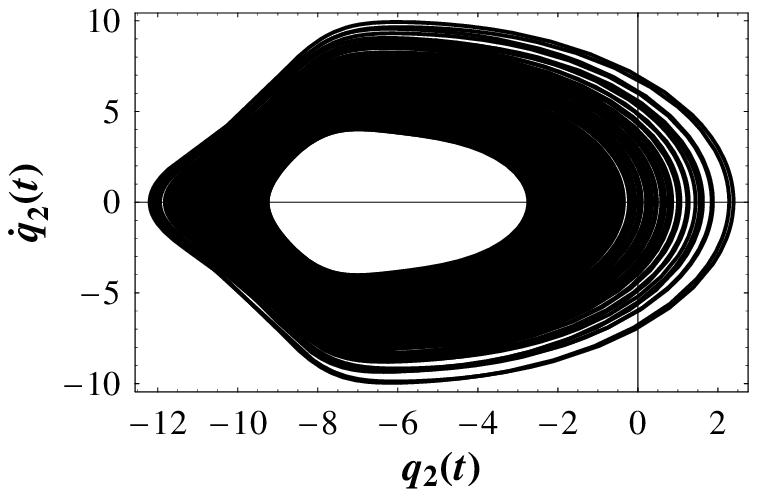}}}
\rotatebox{0.0}{\scalebox{0.72}{\includegraphics{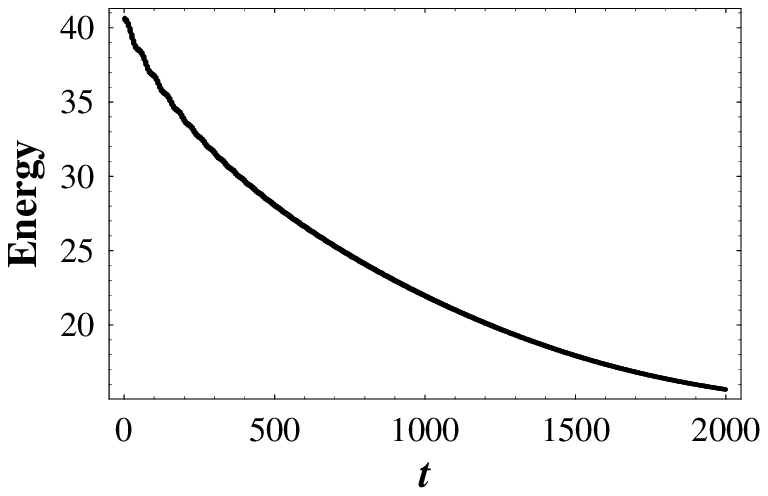}}}
\caption{\footnotesize{Case (II). Phase-space and energy for the inert environment. Oscillator (I) left, oscillator (II) middle. Parameters: $q_{1}=10,\,q_{2}=-10,\,m_{1}=m_{2}=1,\,\omega_{1}=\omega_{2}=1,\, T_{2}=5,\,\mu_{22}=0.3,\,\lambda=0.3$. Initial conditions: $q_{1}(0)=15,\,q_{2}(0)=-10,\,\dot{q}_{1}(0)=\dot{q}_{2}(0)=0$.}}
\label{fig52}\end{figure}

In our second case, the Lagrangian is given by
\begin{eqnarray}
\mathcal{L}\left(\{\dot{q}_{i}\},\,\{q_{i}\}\right)&=&\frac{m_{1}}{2}\,\dot{q}_{1}(t)^{2}\,-\,\frac{m_{1}}{2}\,\omega_{1}^{2}\,\left(q_{1}(t)\,-\,q_{1}\right)^{2}\nonumber\\
&&+\,b_{2}\left[q_{2}(t)\,-\,q_{2}\right]^{-2}\,\left(\frac{m_{2}}{2}\,\dot{q}_{2}(t)^{2}\,-\,\frac{m_{2}}{2}\,\omega_{2}^{2}\,\left(q_{2}(t)\,-\,q_{2}\right)^{2}\right)\,-\,\frac{m_{2}}{4}\,\omega_{2}^{4}\,T_{2}^{2}\,V_{3}^{(2)}[q_{2}(t)\,-\,q_{2}]\,\nonumber\\
&&-\,\frac{\lambda^{2}}{2}\,\left(q_{1}(t)\,-\,q_{2}(t)\right)^{2}\quad.
\label{TIP5}\end{eqnarray}
Now the harmonic potential of the second particle is also immersed into the reservoir. The dynamics is calculated with the same initial conditions as for case (I), the behaviour of the phase-space and the energy are illustrated in Fig. (\ref{fig52}).

When the time-evolution begins, we find the same behaviour as for case (I). Energy flows from oscillator (I) towards oscillator (II), the phase-space of oscillator (II) expands, while the phase-space of oscillator (I) contracts. But, surprisingly, the draining of oscillator (I) continues, and only when it's final state is almost reached, friction takes effect on oscillator (II). From the behaviour of the energy we deduce that after 2000 time-units the system is still far from it's final state, where in case (I) above the final state is already almost reached. The final state in case (II) is only reached when the energy of both oscillators has completely dissipated.

This behaviour may be explained as follows. Oscillator (II) is immersed in a dissipative environment and tries to adopt a state of dynamical balance. The thermal energy of the environment is not at disposal, but the energy of oscillator (I) is available due to the harmonic coupling. The energy of oscillator (I) thereby acts as a reservoir. Consequently, oscillator (II) takes up energy from oscillator (I), but when oscillator (I) is drained the flux towards oscillator (II) slowly breaks down. If so, the dissipation again strongly begins to act on oscillator (II), and it's phase-space starts to contract. One could say that oscillator (II) parasitizes oscillator (I). 

Albeit not as clearly visible as in our present example, the same mechanism works in case (I). However, in case (I) the harmonic potential of oscillator (II) is external, and this obviously leads to a much poorer flux of energy from oscillator (I) towards oscillator (II). Our results for the cases (III/IV) below will support our interpretation about the attempt of oscillator (II) to adopt dynamical balance.

For an external interaction as given by case (I), the behaviour might still be guessed by physical reasoning, however, the case of internal interaction teaches us that expectations very well may be wrong. At a first glance we would not assume that both systems behave so essentially different as they do. This, of course, is also due to the high non-linearity of the equations of motion. Furthermore, the behaviour of the system is extremely sensitive to it's initial conditions and to even small changes of the values of it's parameters.

In conclusion, we learn that external and internal interactions lead to essentially different behaviour of systems. A system completely immersed into a dissipative environment will use any available reservoir to adopt a state of dynamical balance, be it mechanical or thermal.

\subsection{Case III: one-sided back-reaction}
When we allow a back-reaction of oscillator (II) with it's reservoir, the temperature $T_{2}$ becomes time-dependent, and the gyroscopic term has to be included. Thus, the Lagrangian for case (III) reads
\begin{eqnarray}
\mathcal{L}\left(\{\dot{q}_{i}\},\,\{q_{i}\}\right)&=&\frac{m_{1}}{2}\,\dot{q}_{1}(t)^{2}\,-\,\frac{m_{1}}{2}\,\omega_{1}^{2}\,\left(q_{1}(t)\,-\,q_{1}\right)^{2}\nonumber\\
&&+\,b_{2}\left[q_{2}(t)\,-\,q_{2},\,T_{2}(t)\right]^{-2}\,\left(\frac{m_{2}}{2}\,\dot{q}_{2}(t)^{2}\,-\,\frac{m_{2}}{2}\,\omega_{2}^{2}\,\left(q_{2}(t)\,-\,q_{2}\right)^{2}\right)\nonumber\\
&&+\,\dot{q}_{2}(t)\,m_{2}\,\omega_{2}^{2}\,T_{2}(t)\,V_{1}^{(2)}[q_{2}(t)\,-\,q_{2},\,T_{2}(t)]\,-\,\frac{m_{2}}{4}\,\omega_{2}^{4}\,T_{2}(t)^{2}\,V_{3}^{(2)}[q_{2}(t)\,-\,q_{2},\,T_{2}(t)]\nonumber\\
&&-\,\frac{\lambda^{2}}{2}\,\left(q_{1}(t)\,-\,q_{2}(t)\right)^{2}\quad.
\label{TIP6}\end{eqnarray}
The first particle still moves in an ideal and frictionless environment, the second particle is again bound by an internal harmonic potential.

The dynamics of the back-reaction is calculated by $\Delta E_{2}\,=\,-\,\Delta T_{2}$ for each time-step, as above. The energy $E_{2}$ is solely the energy of the second particle without the harmonic coupling.

\begin{figure}[t!]\centering\vspace{0.cm}
\rotatebox{0.0}{\scalebox{0.72}{\includegraphics{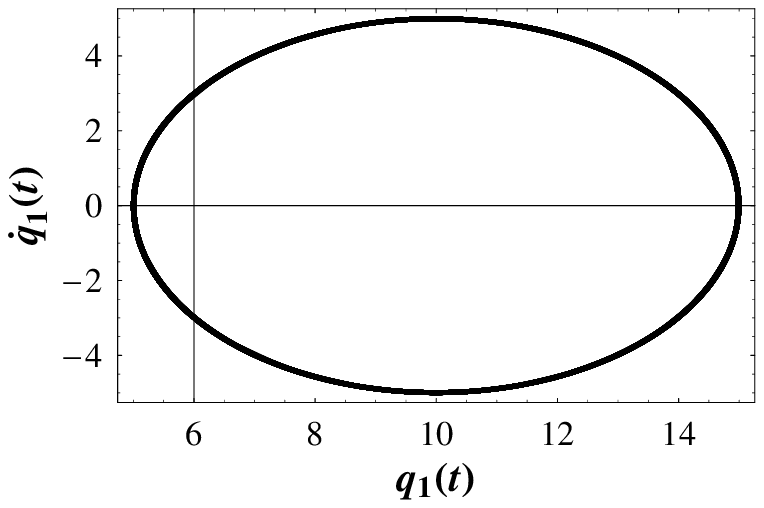}}}
\rotatebox{0.0}{\scalebox{0.72}{\includegraphics{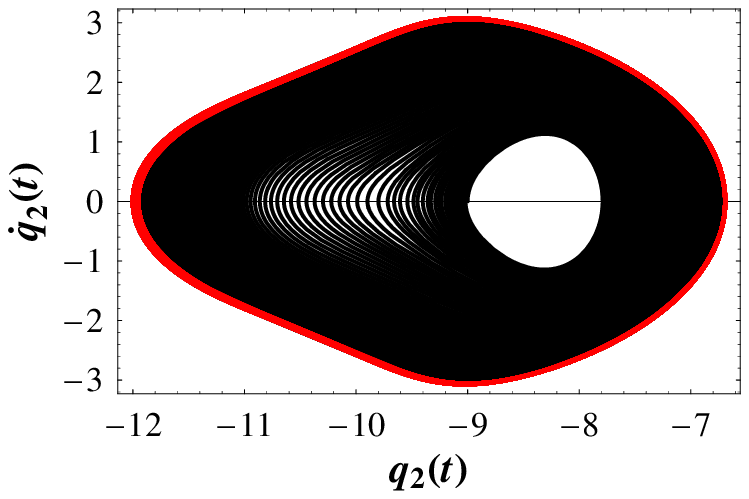}}}
\rotatebox{0.0}{\scalebox{0.73}{\includegraphics{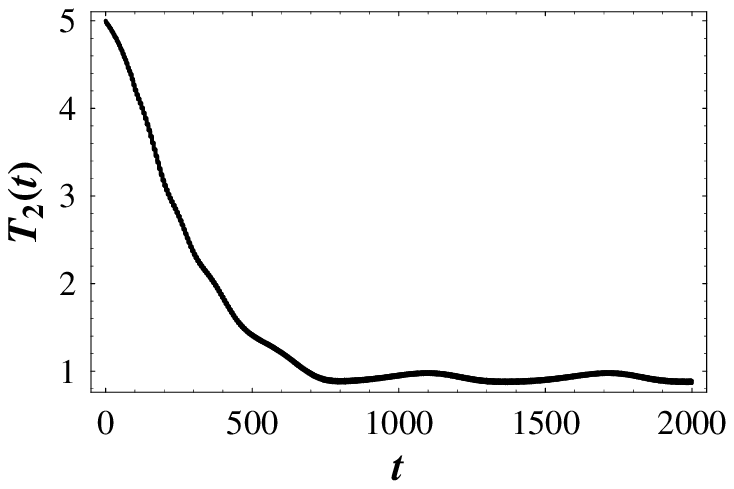}}}
\rotatebox{0.0}{\scalebox{0.73}{\includegraphics{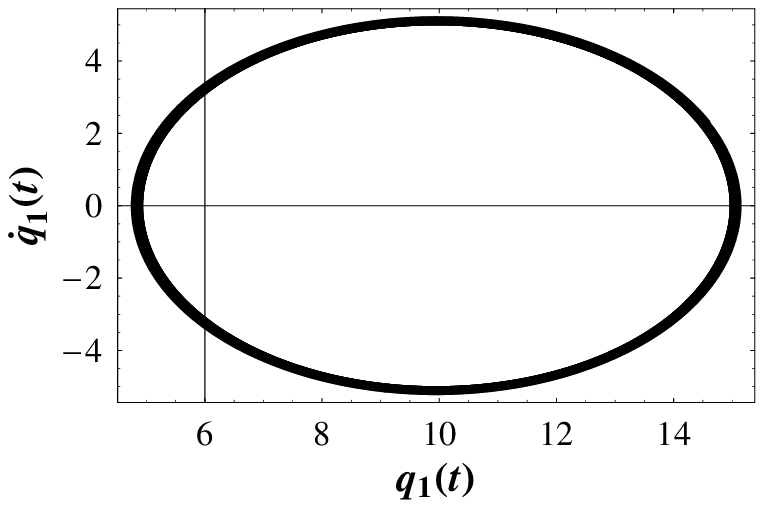}}}
\rotatebox{0.0}{\scalebox{0.72}{\includegraphics{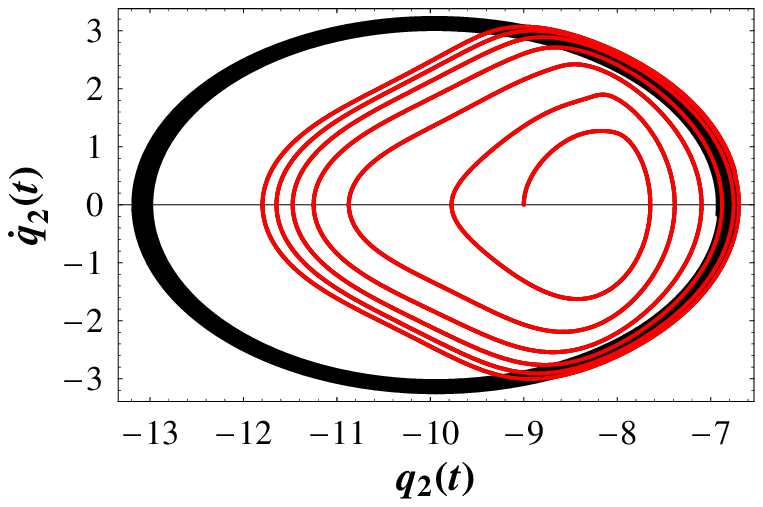}}}
\rotatebox{0.0}{\scalebox{0.72}{\includegraphics{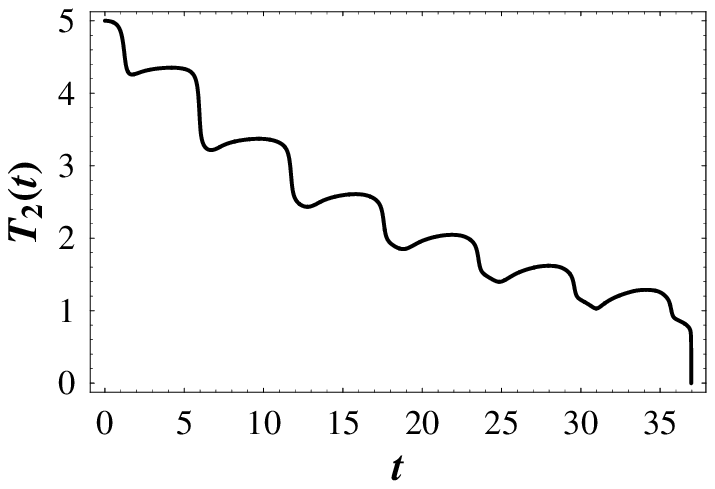}}}
\caption{\footnotesize{Case (III). Phase-space and temperatures. Oscillator (I) left, oscillator (II) middle. Parameters: $m_{1}=m_{2}=1,\,\omega_{1}=\omega_{2}=1,\,T_{2}(0)=5,\, \mu_{22}=0.3,\,q_{1}=10,\,q_{2}=-10,\,q_{1}(0)=15,\,q_{2}(0)=-9,\,\dot{q}_{1}(0)=0,\,\dot{q}_{2}(0)=0$. Upper row: $\lambda=0.01$, the red inlay marks the times $t\in[800,\,2000]$. Lower row: $\lambda=0.05$\,, the red inlay marks the fast expansion of the phase-space while the temperature drops to zero, $t\in[0,\,36.958]$, illustrated for a total time $t\in[0,\,100]$.}}
\label{fig53}\end{figure}

For the analysis of the cases (I/II) we have chosen a coupling-constant $\lambda\,=\,0.3$, in order to get clearly visible result. In our present case (III) even lower couplings $\lambda\,=\,\{0.01,\,0.05\}$ are large enough to discover considerably different dynamical phenomena, see Fig. (\ref{fig53}).

If the thermal energy of the environment is available, oscillator (II), in order to gain dynamical balance, prefers to take up thermal energy instead of draining the energy of oscillator (I). As above, see Fig. (\ref{fig34}), this leads to a cool-down of the environment and an expansion of the phase-space of oscillator (II). However, the influx of thermal energy happens much faster than for the single oscillator. Already for a marginal coupling of $\lambda=0.01$ the influx of thermal energy is at least 2.6-times faster until dynamical balance is reached, when we assign this to the time $t\approx800$, where $T_{2}(t)$ passes through it's first minimum. The red inlay in the upper row of Fig. (\ref{fig53}) marks the time-interval $t\in[800,\,2000]$, where oscillator (II) moves in dynamical balance. We emphasize that oscillator (II) is not only in dynamical balance with it's thermal environment, it is also in dynamical balance with oscillator (I).

In the lower row of Fig. (\ref{fig53}), we show the dynamics for a coupling $\lambda=0.05$. The coupling itself can still be regarded as small, but nevertheless it is five times larger than in the previous case. Here, we find that oscillator (II) literally sucks in the thermal energy of it's environment, and for $t=36.985$, almost 54-times faster than for the single oscillator, the thermal energy even drops to zero. This last transition happens fast but steady, as it can be deduced from the red inlay we have placed in the phase-space of oscillator (II). The red inlay describes the phase-space for $t\in[0,\,36.985]$.

Furthermore, this transition may be understood as a real phase-transition, since for $T_{2}=0$ the system changes essentially. At zero thermal energy the environment freezes out, oscillator (II) completely decouples from it's environment, and because the friction $\mu_{2}$ couples to the thermal energy, the motion becomes non-dissipative. Without further action from the outside, oscillator (II) will remain isolated, and both oscillators behave like two coupled, ideal harmonic oscillators. Thus, oscillator (II) acquires a truly frictionless state.  We chose the term \emph{truly} to distinguish this state from the \emph{apparently} frictionless state, where a system is in dynamical balance with it's thermal environment.

\subsection{Case IV: two-sided back-reaction}
Here, both oscillators can exchange energy with their environments, and thus the Lagrangian is given by
\begin{eqnarray}
\mathcal{L}\left(\{\dot{q}_{i}\},\,\{q_{i}\}\right)&=&\sum_{i=1}^{2}\,b_{i}\left[q_{i}(t)\,-\,q_{i},\,T_{i}(t)\right]^{-2}\,\left(\frac{m_{i}}{2}\,\dot{q}_{i}(t)^{2}\,-\,\frac{m_{i}}{2}\,\omega_{i}^{2}\,\left(q_{i}(t)\,-\,q_{i}\right)^{2}\right)\nonumber\\
&&+\,\dot{q}_{i}(t)\,m_{i}\,\omega_{i}^{2}\,T_{i}(t)\,V_{1}^{(i)}[q_{i}(t)\,-\,q_{i},\,T_{i}(t)]\,-\,\frac{m_{i}}{4}\,\omega_{i}^{4}\,T_{i}(t)^{2}\,V_{3}^{(i)}[q_{i}(t)\,-\,q_{i},\,T_{i}(t)]\nonumber\\
&&-\,\frac{\lambda^{2}}{2}\,\left(q_{1}(t)\,-\,q_{2}(t)\right)^{2}\quad.
\label{TIP7}\end{eqnarray}

\begin{figure}[t!]\centering\vspace{0.cm}
\rotatebox{0.0}{\scalebox{0.88}{\includegraphics{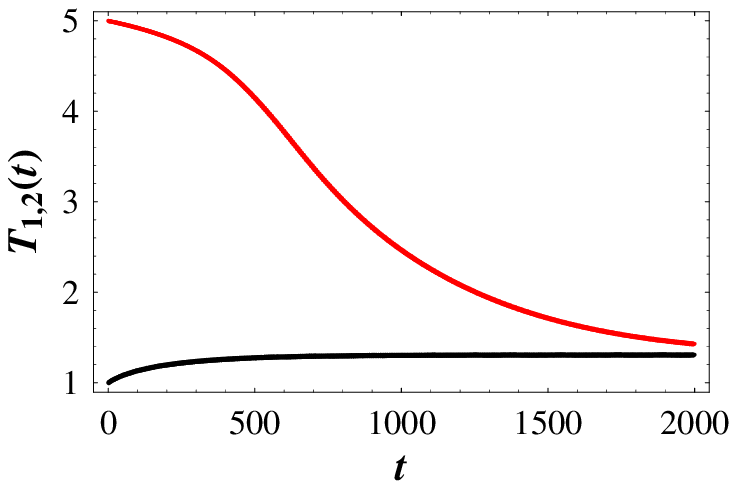}}}
\rotatebox{0.0}{\scalebox{0.85}{\includegraphics{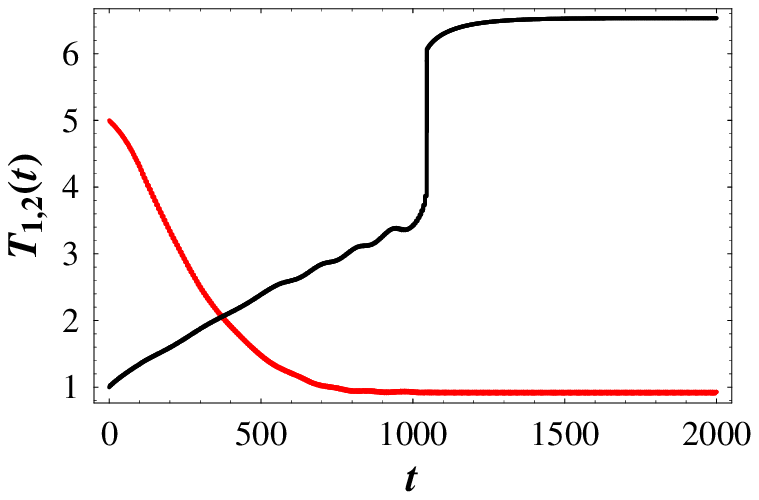}}}
\caption{\footnotesize{Case (IV). Oscillator (I) (black), oscillator (II) (red). Thermal energies of the two-sides back-reaction for $m_{1}=m_{2}=1,\,\omega_{1}=\omega_{2}=1,\,T_{1}(0)=1,\,\mu_{21}=0.3,\,T_{2}(0)=5,\, \mu_{22}=0.3,\,q_{1}=10,\,q_{2}=-10,\,q_{1}(0)=8,\,q_{2}(0)=-9,\,\dot{q}_{1}(0)=0,\,\dot{q}_{2}(0)=0$. Left: $\lambda=0$, right: $\lambda=0.0095$.}}
\label{fig54}\end{figure}

To illustrate the behaviour of a two-sided back-reaction, we chose to couple the two single oscillators depicted in Fig. (\ref{fig34}) with slightly different initial conditions. For better comparison, the behaviour of the thermal energies of the two single oscillators, $\lambda=0$, is shown in Fig. (\ref{fig54}) on the left, while the thermal energies of the coupled oscillators are shown on the right. The black curve describes system (I), the red curve describes system (II). The term system thereby refers to the oscillator and it's environment combined. As for case (III) we calculate the time-evolution of the energies by $\Delta E_{1}=-\Delta T_{1}$ and $\Delta E_{2}=-\Delta T_{2}$. Note that there is only a mechanical coupling between the systems, but thermally they are still isolated.

We deduce that already for an apparently marginal coupling $\lambda=0.0095$ the systems behave essentially different, compared to the non-coupled case. Oscillator (I) gains dynamical balance by a considerably large transfer of energy towards it's environment, while oscillator (II) absorbs thermal energy from it's environment about 4-times faster than in the uncoupled case. The total exchange of thermal energy is not equal, since $\left|\Delta T_{1}^{\rm{(tot)}}\right|>\left|\Delta T_{2}^{\rm{(tot)}}\right|$, however, the internal energy,
\begin{equation}
U_{i}\,=\,E_{i}(0)\,+\,T_{i}(0)\,=\,E_{i}(t)\,+\,T_{i}(t)\quad,
\label{TIP8}\end{equation}
of both systems is of course conserved. The steep transition of system (I) towards it's final state is continuous.

The flux of energy through the harmonic coupling is almost negligible, but our results show that already a marginal perturbation from the outside leads to essential changes of the dynamics of interacting systems in a reactive thermal environment. We assume that both oscillators seek to acquire dynamical balance with respect to their thermal environment, with respect to each other, and with respect to the harmonic coupling. A state of dynamical balance also with respect to the harmonic coupling seems to be reasonable, since else the essential reaction on the marginal coupling $\lambda=0.0095$ remains inexplicable.

Due to the high non-linearity of the equation of motion, the details of the dynamics are generally unpredictable. The only hint about the likely behaviour in our present example is the initial difference of the thermal energies $T_{2}(0)>T_{1}(0)$. However, there are also scenarios for different values of $\lambda$ where $T_{2}$ still grows, while $T_{1}$ drops. In conclusion, we find that a reactive thermal environment is extremely sensitive to a perturbation, and that the details of the behaviour due to the high non-linearity is only hardly predictable by physical reasoning. This sensitivity of course also includes the initial conditions of the time-evolution. As already discussed for the case of a single oscillator, the initial conditions decide about the principal stability of the system.

\section{Heat and entropy}
When we suppose that a system and it's environment is closed, the law of energy-conservation can be set up as
\begin{equation}
Q(t)\,+\,E(t)\,+\,T\,=\,U_{0}\quad,
\label{Heat1}\end{equation}
where $Q$ is the heat, $E$ is the energy of the system, $T$ the thermal energy of the environment, which also could be dropped here, and $U_{0}$ is the conserved internal energy. From Eq. (\ref{Heat1}) it follows
\begin{equation}
\dot{Q}(t)\,=\,-\,\dot{E}(t)\quad,
\label{Heat2}\end{equation}
which implies that dissipated energy generates heat as it must, since in a dissipative system $\dot{E}(t)<0$ holds. Thus, all examples with an inert thermal environment we have treated above generate heat, as it was to be expected. However, we have to be a little more precise here. Our distinction between heat $Q$ and thermal energy $T$ comes from our assumption that the thermal reservoir the oscillator moves in is inert. This means that all heat produced dissipates out of the oscillator, and does not increase the thermal energy of the environment. If else, any heat produced increases the thermal energy of the environment and thus leads to a modulation of the dynamics of the oscillator. For the numerical treatment this means that the numerical value of the thermal environment must be increased after each time-step, $T_{i+1}=T_{0}+Q_{i}$. We emphasize that such a scenario must not be confused with the case of a reactive thermal environment, where both directions are open for the flux of energy, such that dynamical balance can develop. In our present case we assume that energy can only flow into one direction, that is away from the oscillator.

By the relation between heat and entropy, $Q(t)=\Theta(t)\,S(t)$, where $\Theta(t)$ is the temperature, we can rewrite Eq. (\ref{Heat2}) and obtain
\begin{equation}
\dot{\Theta}(t)\,S(t)\,+\,\Theta(t)\,\dot{S}(t)\,=\,-\,\dot{E}(t)\quad,
\label{Heat3}\end{equation}
which yields
\begin{equation}
\dot{S}(t)\,=\,-\,S_{0}\,\Theta_{0}\,\frac{\dot{\Theta}(t)}{\Theta(t)^{2}}\,-\,\frac{\dot{E}(t)}{\Theta(t)}\,+\,\frac{E(t)}{\Theta(t)^{2}}\,\dot{\Theta}(t)\quad.
\label{Heat4}\end{equation}
Reintroducing our definition of the thermal energy $T$, see Eq. (\ref{FFO1}), we cast Eq. (\ref{Heat4}) into
\begin{equation}
\dot{S}(t)\,=\,-\,S_{0}\,T_{0}\,\frac{\dot{T}(t)}{T(t)^{2}}\,-\,\frac{2\,k_{\rm{B}}}{m\,\omega^{2}}\,\left(\frac{\dot{E}(t)}{T(t)}\,-\,\frac{E(t)}{T(t)^{2}}\,\dot{T}(t)\right)\quad.
\label{Heat5}\end{equation}

If the thermal environment is inert, thus $T=\rm{const.}$, we immediately see that the growth of the entropy is negligible, since then holds
\begin{equation}
\dot{S}(t)\,=\,-\,\frac{2\,k_{\rm{B}}}{m\,\omega^{2}}\,\frac{\dot{E}(t)}{T}\quad.
\label{Heat6}\end{equation}
Boltzmann's constant $k_{\rm{B}}$ is small, and a considerable change is only noticeable for a real many-body system
\begin{equation}
\dot{S}(t)\,=\,-\,\frac{2\,k_{\rm{B}}}{T}\,\sum_{n=1}^{N_{A}}\frac{\dot{E}_{n}(t)}{m_{n}\,\omega_{n}^{2}}\quad,
\label{Heat7}\end{equation}
where $N_{A}$ is Loschmidt's number.

If the thermal environment is reactive, the situation looks completely different. For our examples above, the main contribution then comes from
\begin{equation}
\dot{S}(t)\,=\,-\,S_{0}\,T_{0}\,\frac{\dot{T}(t)}{T(t)^{2}}\quad,
\label{Heat8}\end{equation}
given that $S_{0}$ is already finite. As it holds for $\dot{Q}(t)$, Eq. (\ref{Heat8}) tells us that $\dot{S}(t)>0$ if $\dot{T}(t)<0$. Since time-dependent thermal energies of oscillators have tiny but extremely rapid oscillations, it is also convenient to consider the time average,
\begin{equation}
\left<\dot{S}(t)\right>\,=\,\frac{1}{t_{\rm{max}}}\,\int_{0}^{t_{\rm{max}}}\,d\tau\,\dot{S}(\tau)\quad.
\label{Heat9}\end{equation}

\begin{figure}[t!]\centering\vspace{0.cm}
\rotatebox{0.0}{\scalebox{0.88}{\includegraphics{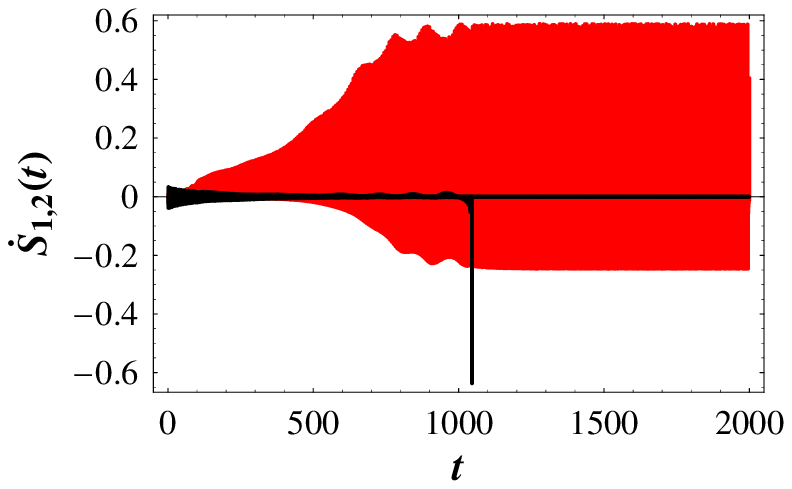}}}
\rotatebox{0.0}{\scalebox{0.92}{\includegraphics{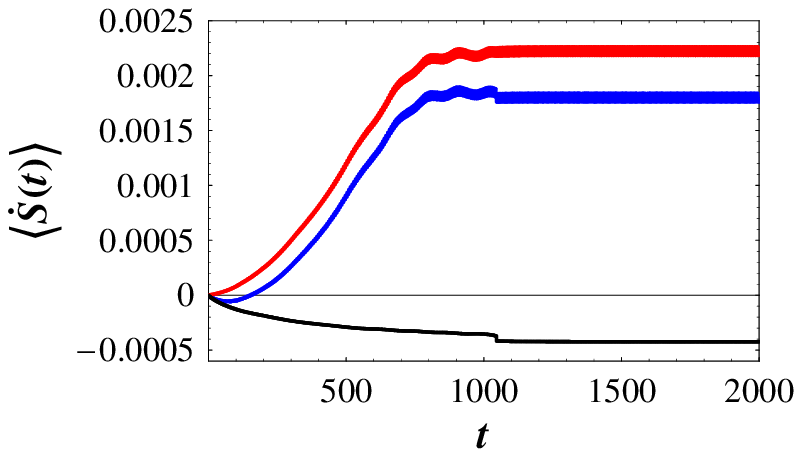}}}
\caption{\footnotesize{Entropies of case (IV) as illustrated in Fig. (\ref{fig54}) for $S_{0}=1$. Left: entropies according to Eq. (\ref{Heat8}. Right: time-averages of the entropies. Blue curve: total averaged entropy.}}
\label{fig55}\end{figure}

In Fig. (\ref{fig55}) we illustrate the behaviour of the entropies of case (IV) according to Eqs. (\ref{Heat8}, \ref{Heat9}), with the initial condition $S_{0}=1$. On the left, we see the impact of the rapid oscillations of $\{T_{1}(t),\,T_{2}(t)\}$. In the averages on the right, the oscillations are damped away, and it turns out that the effective growth of the entropies is actually smaller by orders of magnitude. The entropy of oscillator (I), black curves, declines, since $\dot{T}_{1}(t)>0$, but the behaviour of the total entropy of the system still obeys $\left<\dot{S}(t)\right>>0$. Thus, we understand that the effect of dynamical balance can lead to a decline of the entropy of a sub-system. Furthermore, if the initial condition for the entropy $S_{0}$ is smaller or zero, this effect diminishes or vanishes completely. As can be deduced from the right side of Fig. (\ref{fig55}), the effective dynamics of the entropy is still small enough to be negligible. In conclusion, we say that systems like our chosen examples show entropic dynamics, but still small enough to be regarded as negligible.

Much more interesting than the negligible dynamics of the entropy is the fact that systems like case (IV) do neither produce nor absorb heat. The conservation of energy is given by
\begin{equation}
Q(t)\,+\,E(t)\,+\,T(t)\,=\,U\quad\Rightarrow\quad \dot{Q}(t)\,+\,\dot{E}(t)\,+\,\dot{T}(t)\,=\,0\,\quad\Rightarrow\quad \dot{Q}(t)\,=\,0\quad,
\label{Heat10}\end{equation}
since $\dot{E}(t)=-\,\dot{T}(t)$ holds. The same can be obtained from Eq. (\ref{Heat8}) by direct integration
\begin{equation}
T(t)\,S(t)\,=\,S_{0}\,T_{0}\quad.
\label{Heat11}\end{equation}
Consequently, the dynamics of the thermal energy and the entropy is an internal adjustment, which balances each other, such that heat is conserved in the sense that no heat dissipates out of the whole system. Hence, dynamics like the present one seem to belong to the regime of adiabatic processes. Furthermore, this means that an adiabatic process is the thermodynamic analogon to the mechanical phenomenon of dynamical balance. Since dynamical balance buffers friction, and friction is responsible for the production of heat, this analogon appears to be reasonable. When we identify the state of dynamical balance of a small system with the equilibrium-state of a large system, then the analogon becomes even more reasonable. Remind that adiabaticity only holds for oscillator and environment combined. Oscillator and environment taken alone still heat up or cool down in the sense of absorbing or releasing energy.

\section{Conclusion}
The work we present here explores the range and the applicability of the Onsager-Machlup theory for Lagrangian dynamics in inhomogeneous and thermal environments.

In the introduction, we discussed the place of the Onsager-Machlup Lagrangian amongst Lagrangian approaches to dissipation and friction. We concluded that the Onsager-Machlup approach very likely is the most general and systematic one. This especially may be confirmed by it's relation to the Helmholtz-factor, and thus the Jacobi-multiplier. Furthermore, we have mentioned that the Caldirola-Kanai Hamiltonian belongs to the Onsager-Machlup family, and thus also does Bateman's Lagrangian in a non-obvious way. In this context, other seemingly different Lagrangians may exist, which can be traced back to the Onsager-Machlup Lagrangian.

We have shown that the Fokker-Planck equation and the Onsager-Machlup Lagrangian are related by an easy and straight-forward calculation, if the operator-ordering of the Fokker-Planck equation is taken care of. As to our knowledge, this route of derivation has not been walked elsewhere so far.

Our route of derivation provided us with arguments for an objection against current opinions about the valid terms in the Onsager-Machlup Lagrangian. The common opinion is that some terms in our version and Dekker's version \cite{Dekker1} should not be present, since the transition-amplitude only then is invariant under stochastic coordinate-transform. Our objection is based on purely physical grounds. We say that neither the Fokker-Planck equation, nor the Onsager-Machlup theory are in need of stochastic calculus to have a meaning. Furthermore, the Fokker-Planck equation and thus the Onsager-Machlup theory initially have not been related to stochastic calculus, only Itô's lemma establishes this link. To our regards, the reverse direction is true. Stochastic calculus needs the connection to the Fokker-Planck equation and thus the Onsager-Machlup theory to assign a beyond phenomenological meaning to the Langevin-equation. Thus, any problems with stochastic coordinate-transform do neither affect the structure of the Fokker-Planck equation, nor the Onsager-Machlup Lagrangian. Any attempts to modify the Onsager-Machlup Lagrangian strike back on the Fokker-Planck equation, as our derivation shows. Consequently, the problem of invariance is a purely mathematical problem with no outreach to physics, and thus it must be resolved on it's own grounds.

After transforming the Onsager-Machlup Lagrangian into it's single-particle version by using Ornstein's fluctuation-dissipation theorem, we gave arguments about how interaction can be included. We distinguished between internal and external interaction. By internal interaction, we mean the interaction of a system, which is completely immersed into it's environment. By external interaction, we mean an interaction, which acts upon the immersed system from outside of the environment. Furthermore, we gave arguments about the general structure of the environment.

By inspecting the equation of motion, we found that the Onsager-Machlup Lagrangian generates Stokes-friction and Newtonian friction in a natural way. This provided us with hints of how the environmental function $b[q(t)]$ can be determined.

Our first example is the free fall through a barometric medium. We started with the well-known phenomenological equation of motion with Newtonian friction, and after the calculation of the environmental function $b[q(t)]$ by the Newtonian friction term, we determined the full Onsager-Machlup Lagrangian. As a first result, we found that deviations from the phenomenological behaviour only occur for very small masses or high temperatures. This insinuates that the resistivity of a barometric medium against directional motion in these cases is higher as it is to be expected by the phenomenological approach. Furthermore, we modeled a blow of wind by a solitonic pulse. Our discussion of the physically reasonable structure of the environmental function $b[q(t)]$ in this case confirmed our arguments about the general structure of the function $b[q(t)]$. The trajectories we have calculated for a blow of wind are reasonable, and thus we may conclude that Onsager-Machlup dynamics can model systems on the base of classical mechanics, which else are not accessible to classical mechanics.

As a second example, we treated a single-particle oscillator evolving in a thermal environment of classical oscillators. For an inert thermal environment we found dissipative behaviour as it was to be expected. Furthermore, we allowed the thermal environment be reactive. As we understand it, the allowance of back-reaction yielded one of the two major results of this paper: the flux of energy between mechanical system and thermal environment generates a state of dynamical balance. Dynamical balance buffers friction, and so the mechanical system apparently evolves like an ideal oscillator without friction. In terms of macroscopic systems dynamical balance has the same meaning as an equilibrium. Moreover, the behaviour of the time-dependent thermal energy $T(t)$ shows tiny but extremely rapid oscillations, which may be interpreted as fluctuations around a stable mean-value. In this again we find an analogon to a thermodynamic equilibrium, where also small fluctuations around the mean-value are present.

As a third example, we treated two harmonically coupled oscillators in four variations. In the first three variations oscillator (I) is always an ideal one, in the fourth variation also oscillator (I) is immersed in a thermal environment.

In case (I) only the particle of oscillator (II) is immersed into the environment, it's confining harmonic potential is external. We found that the ideal oscillator (I) is only marginally affected by this configuration, while oscillator (II) shows dissipative behaviour, as it was to be expected.

In case (II) this changes essentially. Here oscillator (II) is fully immersed, such that it's interaction is internal. Oscillator (II) tries to adopt a state of dynamical balance. As oscillator (II) has no access to it's thermal environment, it takes oscillator (I) as a reservoir of energy and drains it. Only after oscillator (I) is almost completely drained friction starts to take effect on oscillator (II). This shows that a harmonic coupling is not suitable for a state of dynamical balance, it only slows down the effect of friction oscillator (II) is subject to. However, we conclude that this behaviour is not what was to be expected. It shows that there indeed is a crucial difference between internal and external interaction.

In case (III) oscillator (II) has access to it's thermal environment. As a result, we observed that now oscillator (II) prefers to take up energy from it's environment instead of draining oscillator (I). A state of dynamical balance is finally acquired. In our second choice for the numerical value of the strength of the harmonic coupling, we even observed a real phase-transition. Oscillator (II) drains it's thermal environment completely and thus begins to evolve like an ideal oscillator. Since then the effect of friction is gone, this transition is irreversible without any further action from the outside. From this we learn that a thermal reservoir is always the preferred source of energy compared to a mechanical reservoir of energy.

In case (IV) finally both oscillators are fully immersed in a reactive thermal environment. Only a mechanical coupling is present, thermally both systems are still isolated. We observed that the state of dynamical balance is reached relatively fast. The more interesting result is that already a marginal mechanical coupling between the two oscillators is enough to create a behaviour, that is essentially different from the case of a zero mechanical coupling. Thus, a system in a reactive thermal environment is extremely sensitive to perturbations from the outside. From this we conclude that the oscillators do not only seek dynamical balance with respect to each other and their thermal environments, but also with respect to the perturbation.

As a last point, we treated the production of heat and entropy. In the case of an inert thermal environment heat is produced due to the action of friction and dissipates out of the system. This was to be expected. For the case of dynamical balance but we found behaviour, that may be interpreted as the analogon of an adiabatic process. No heat is produced, the aim of the oscillators to adopt a state of dynamical balance leads to an internal adjustment between entropy and thermal energy, such that the internal energy of the system is conserved.

The first major result of this paper certainly is the method of how the environmental function $b[q(t)]$ can be determined for thermal environments. As second major result of this paper we regard the aim of completely immersed systems to acquire a state of dynamical balance if a reservoir of energy is accessible. Dynamical balance buffers the action of friction and thus may be understood as a mechanical analogon to the state of the thermodynamic equilibrium of large systems. Furthermore, the behaviour of the entropy in this case insinuates that dynamical balance is also related to adiabatic processes, which makes the analogy even more reasonable.

Putting all this together, we carefully draw the conclusion that single-particle dynamics by the Onsager-Machlup Lagrangian can simulate the behaviour of thermodynamic systems, either in the non-equilibrium or the equilibrium regime.

An outline about further work is difficult, since the possibilities of the Onsager-Machlup theory are rich. So far, we refer to the sequel to this paper, where we discuss how the Newtonian equation of motion with stochastic noise correctly should be treated by the Onsager-Machlup theory. The result is a Lagrangian, which describes all possible actions of environments all in one. Furthermore, we provide rules about how the hierarchy of environments and actions from the outside must be included into the Lagrangian.

\end{document}